\documentclass[epj]{svjour}
\usepackage{amssymb}
\usepackage{graphics}
\begin{document}
\title{The $pK^0\Sigma^+$ final state in proton-proton collisions}

\subtitle{{\it The  COSY-TOF collaboration}}

\mail{\mbox{m.schulte-wissermann@physik.tu-dresden.de}  (M.~Schulte-Wissermann)}

\author{M.~Abdel-Bary\inst{3}   \and
S.~Abdel-Samad\inst{3}   \and
K-Th.~Brinkmann\inst{1}$^{,}$ \inst{8}   \and
H.~Clement\inst{4}   \and
J.~Dietrich\inst{1}   \and
E.~Doroshkevich\inst{4}   \and
S.~Dshemuchadse\inst{1}   \and
K.~Ehrhardt\inst{4}   \and
A.~Erhardt\inst{4}   \and
W.~Eyrich\inst{2}   \and
A.~Filippi\inst{7}   \and
H.~Freiesleben\inst{1}   \and
M.~Fritsch\inst{2}   \and
W.~Gast\inst{3}   \and
J.~Georgi\inst{2}   \and
A.~Gillitzer\inst{3}   \and
J.~Gottwald\inst{1}   \and
D.~Hesselbarth\inst{3}   \and
H.~J\"ager\inst{3}   \and
B.~Jakob\inst{1}   \and
R.~J\"akel\inst{1}   \and
L.~Karsch\inst{1}   \and
K.~Kilian\inst{3}   \and
H.~Koch\inst{9}   \and
M.~Krapp\inst{2}   \and
J.~Kre\ss\inst{4}   \and
E.~Kuhlmann\inst{1}   \and
A.~Lehmann\inst{2}   \and
S.~Marcello\inst{7}   \and
S.~Marwinski\inst{3}   \and
S.~Mauro\inst{9}   \and
W.~Meyer\inst{9}   \and
P.~Michel\inst{5}   \and
K.~M\"oller\inst{5}   \and
H.~P.~Morsch\inst{6}   \and
H.~M\"ortel\inst{2}   \and
L.~Naumann\inst{5}   \and
N.~Paul\inst{3}   \and
L.~Pinna\inst{2}   \and
C.~Pizzolotto\inst{2}   \and
Ch.~Plettner\inst{1}   \and
S.~Reimann\inst{1}   \and
M.~Richter\inst{1}   \and
J.~Ritman\inst{3}   \and
E.~Roderburg\inst{3}   \and
A.~Schamlott\inst{5}   \and
P.~Sch\"onmeier\inst{1}   \and
W.~Schroeder\inst{2}$^{,}$\inst{3} \and
M.~Schulte-Wissermann\inst{1}   \and
T.~Sefzick\inst{3}   \and
F.~Stinzing\inst{2}   \and
M.~Steinke\inst{9}   \and
G.~Y.~Sun\inst{1}   \and
A.~Teufel\inst{2}   \and
W.~Ullrich\inst{1}   \and
G.~J.~Wagner\inst{4}   \and
M.~Wagner\inst{2}   \and
R.~Wenzel\inst{1}   \and
A.~Wilms\inst{9}   \and
P.~Wintz\inst{3}   \and
S.~Wirth\inst{2}   \and
P.~W\"ustner\inst{3}   \and 
P.~Zupranski\inst{6}}

%\author{First author\inst{1} \and Second author\inst{2}% etc
%% \thanks is optional - remove next line if not needed
%}                     % Do not remove
%
%\offprints{m.schulte-wissermann@physik.tu-dresden.deoderpostadressefragezeichen}          % Insert a name or remove this line
%
%\institute{Insert the first address here \and the second here}
\institute{Institut f\"ur Kern- und Teilchenphysik, Technische Universit\"at Dresden, D-01062 Dresden, Germany \and
		   Physikalisches Institut, Universit\"at Erlangen-N\"urnberg, D-91058 Erlangen, Germany \and
		   Institut f\"ur Kernphysik, Forschungszentrum J\"ulich, D-52425 J\"ulich, Germany \and
		   Physikalisches Institut, Universit\"at T\"ubingen, D-72076 T\"ubingen, Germany \and
		   Institut für Strahlenphysik, Helmholtzzentrum Dresden-Rossendorf, Postfach 51 01 19, D-01314 Dresden, Germany \and
		   Soltan Institute for Nuclear Studies, 05-400 Swierk/Otwock, Poland \and
		   INFN Torino, 10125 Torino, Italy \and
		   Helmholtz Institut f\"ur Strahlen- und Kernphysik, Rheinische Friedrich-Wilhelm-Universit\"at Bonn, D-53115 Bonn, Germany \and
		   Institut f\"ur Experimentalphysik, Ruhr-Universit\"at Bochum, D-44780 Bochum, Germany}
\date{Received: date / Revised version: date}

\abstract{ This paper reports results from a study of the reaction $\mathrm{pp\to pK^0\Sigma^+}$ at beam momenta
of $p_{beam}$ = 2950, 3059, and 3200 MeV/c (excess energies of $\epsilon=$ 126, 161, and 206 MeV). 
Total cross sections were determined for all energies; a set of differential cross sections (Dalitz plots; invariant mass 
spectra of all two-body subsystems; angular distributions of all final state particles; distributions in helicity 
and Jackson frames) are presented for $\epsilon=$ 161 MeV. The total cross sections 
are proportional to the volume of available three-body phase-space 
indicating that the transition matrix element does not change significantly in this range 
of excess energies. It is concluded from the differential data that the reaction
proceeds dominantly via the $N(1710)P_{11}$ and/or $N(1720)P_{13}$ resonance(s);
$N(1650)S_{11}$ and $\Delta(1600)P_{33}$ could also contribute.
}

 % end of PACS codes

\maketitle

\section{Introduction}
\label{intro}

The study of associated strangeness production in pro\-ton-proton collisions is one of the 
major physics programs carried out at the COoler SYn\-chro\-tron COSY located at 
For\-schungs\-zen\-trum J\"ulich, Germany. Various experimental groups have contributed
data to the final states  $pK^+\Lambda$, $pK^+\Sigma^0$, and $nK^+\Sigma^+$ in the past decade
\cite{grzonka97,balewski97,balewski98,sewerin99,COSY11kowina04,toflambda06,COSY-11nks,valdau07,toflambda10,ddlambdasigma10,valdau10,HIRES10}. 
In all these three cases the final states contain a rather long-lived charged kaon ($c\tau_{K^+}=3.7\,m$)
and a nucleon, and thus are experimentally well suited to be accessed in inclusive and 
exclusive measurements at the COSY facility. 
In the case of $\Lambda$ and $\Sigma^0$ production 
this effort has led to excitation functions measured at excess energies ($\epsilon=\sqrt s-(m_p+m_K+m_Y)$) 
from only a few MeV above the threshold to $\epsilon \approx 250\,\mathrm{MeV}$.
Dalitz plots were investigated by the COSY-TOF collaboration \cite{toflambda06,toflambda10} which,   
very recently, also published differential cross section data for both reaction 
channels \cite{ddlambdasigma10}. Along with the wealth of data, various theoretical approaches 
based on very different footings were proposed  \cite{Sib2006a,Sib2006b,luizou06,Gasparin,Dillig,Laget91,Laget01}.
In the case of the $nK^+\Sigma^+$ final state total cross sections were published for $\epsilon<200\,\mathrm{MeV}$
\cite{COSY-11nks,valdau07,valdau10,HIRES10} which, however, are contradicting each other strongly.
This oddity is subject to an ongoing theoretical discussion \cite{valdauwilkins10}.

Compared to the three final states discussed above the experimental data basis for the reaction $pp\to pK^0\Sigma^+$ 
is very scarce for $\epsilon \le$ 350 MeV. 
In fact, the only experimental data for this reaction stem from the search for the supposed
pentaquark state $\Theta^+$ ($pp\to \Sigma^+\Theta^+, \Theta^+\to pK^0$) 
carried out by the COSY-TOF collaboration \cite{tof07}.
At higher excess energies measurements of total cross sections were published originating from the 
bubble chamber experiments in the 1960s 
\cite{Louttit61,bierman66,alexander67,Holmgren68,sondhi68,chinowsky68,Firebaugh68,Dunwoodie68} (compiled in \cite{baldini88}).
These data show, although with a large scatter, the total cross section to be rather constant ($\approx 20\,\mu b$) over 
a wide energy range ($350\,\mathrm{MeV} < \epsilon < 2000\,\mathrm{MeV}$).
One of these early experiments reports a Dalitz plot \cite{bierman66} for $\epsilon=723\,\mathrm{MeV}$. 
The approximately 30 entries show an enhanced density at small $K^0\Sigma^+$ invariant masses which was
interpreted as a hint for pion exchange. 

\begin{figure}[ttt]
\resizebox{.5\textwidth}{!}{ \includegraphics{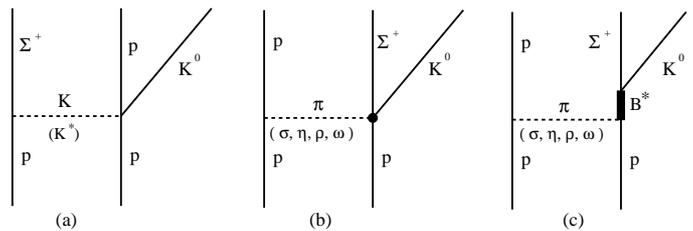} }

\caption{Reaction mechanisms involving (a) strange, (b) non-resonant non-strange, and (c) resonant, non-strange meson exchange 
for $\mathrm{pp\to pK^0\Sigma^+}$. 
Initial and final state interactions are not indicated.}
\label{fig:reactionmechnism}  
\end{figure}

Also from the theoretical point of view, the reaction $pp\to pK^0\Sigma^+$ was only poorly
addressed as compared to the reaction channels containing a $K^+$ meson. 
In 1960, Ferrari \cite{ferrari60} was the first to predict the total cross section for excess energies
between 72 and 400 MeV 
in a meson exchange model where pion as well as kaon exchange were considered. 
A schematic diagram of the respective exchange graphs is shown in fig.~\ref{fig:reactionmechnism} on the left for
kaon exchange (strangeness manifests itself in the exchanged boson) and in the middle for 
pion exchange (associated strangeness production at the $p\pi\to K^0\Sigma^+$ vertex). 
Another early calculation of the cross sections simultaneously for all $pp\to NKY$ channels was performed
by Tsu Yao \cite{tsuyao62} 
for a beam energy of $T_p =2.85$ GeV. This model is based solely on single pion exchange
and the calculated total cross section for $pp\to pK^0\Sigma^+$ ($\epsilon=353\,\mathrm{MeV}$)  is in 
rather good agreement with the prediction of Ferrari \cite{ferrari60} and the only experimental value 
\cite{Louttit61} known at that time. Later on, theoretical analyses and interpretations appeared mainly in experimental papers
 \cite{bierman66,alexander67,sondhi68,chinowsky68}. 
In 1968, Ferrari and Serio \cite{ferrari68} explained fairly well in a meson exchange model 
all the then known total and differential cross sections for the NKY final state by introducing 
empirical cutoff factors in order to model various damping effects at large momentum transfer 
(form factors, absorption). 
Since then, the theoretical progress in describing the production of the $pK^0\Sigma^+$ final state was strongly hampered 
by the lack of new experimental data. 
In view of the experimental and theoretical progress made for the 
$pK^+\Lambda$ and $pK^+\Sigma^0$ final states, in particular data at lower excess energies 
are highly desirable.
This would render possible the further development of more recent theoretical 
approaches \cite{dover93,sibirtsev99,tsuschima99}, which now also include, 
apart from kaon and pion exchange, a production scenario involving intermediate 
baryon resonances ($pp\to B^*p;~B^*=N^*, \Delta^*;~B^*\to K^0\Sigma^+$, 
see fig.~\ref{fig:reactionmechnism}c). If it is assumed 
that the associated strangeness production in pp-reactions procedes via 
resonances, a comprehensive study needs the consideration of $N^*$ as well as 
$\Delta^{+*}$ resonances in the case of both $pp \to pK^+\Sigma^0$ and 
$pp \to pK^0\Sigma^+$ reactions; due to isospin conservation the reaction 
$pp \to pK^+\Lambda$ can involve only $N^*$ resonances. The 
$pp \to nK^+\Sigma^+$-reaction, however, can proceed only via $\Delta^{++*}$-resonances.

In this paper we report results obtained for the reaction $pp\to pK^0\Sigma^+$ measured 
for excess energies of $\epsilon= 126, 161, 206\,\mathrm{MeV}$ ($p_{beam} = 2950, 3059$, and $3200\,\mathrm{MeV/c}$). 
These data supplement earlier studies on the reaction channels 
$pp\to pp\omega$  \cite{mswpaper06,wu09}, $pp\to pK^+\Lambda$, and $pp\to pK^+\Sigma^0$ \cite{ddlambdasigma10}. 
As the data for these various reactions were taken simultaneously and the analyses utilized 
the same software package \cite{MSWdiss} the results obtained are characterized by a very 
high degree of internal consistency. This is, in particular, important for the future development of 
theoretic models when the three channels with associated strangeness production are to be described simultaneously. 

For all three excess energies total cross sections will be given. 
For the data taken at $\epsilon = 161\,\mathrm{MeV}$ Dalitz plots, invariant mass spectra, 
distributions in the CMS, in helicity as well as in Jackson frames will be presented.
All differential distributions are scrutinized with the aim of gaining 
insight into the reaction mechanism. It will be presented that all differential distributions 
can be described simultaneously and consistently if a 
resonant production of the final state is assumed.

%%-----------------------------------------------------------------

%-----------------------------------------------------------------
\begin{figure}[tth]
\resizebox{0.5\textwidth}{!}{ \includegraphics{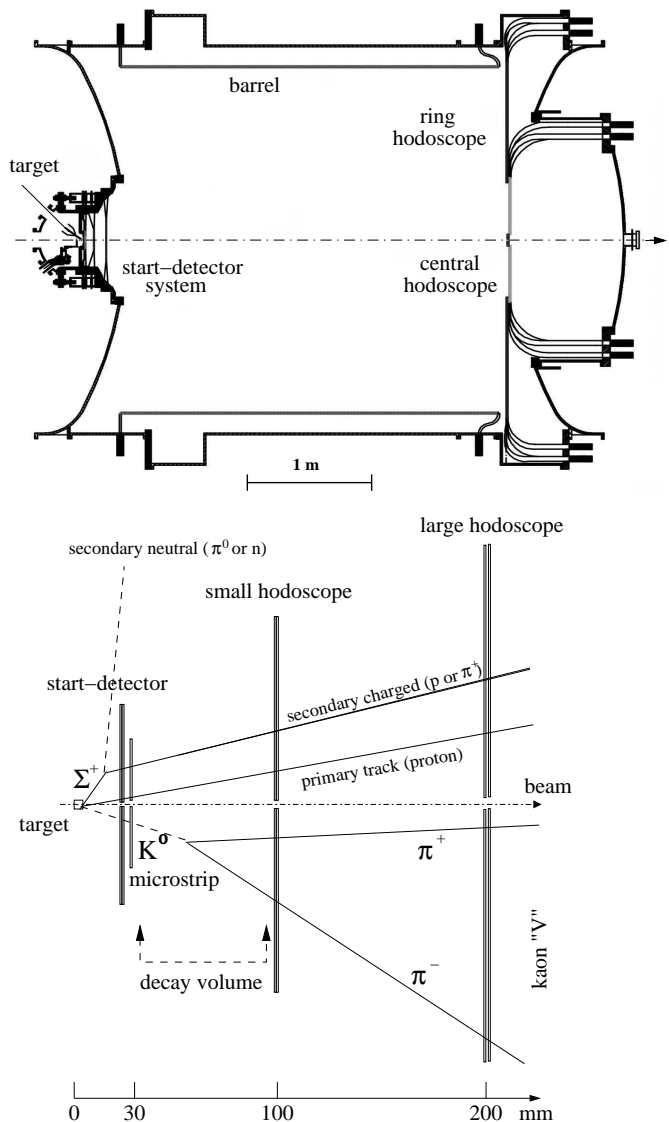} }
\caption{The {\it COSY-TOF} detector (top), the near-target region (start-detector system, bottom).
The lower picture shows a typical event pattern for the $\mathrm{pK^0\Sigma^+}$ exit channel.
}
\label{fig:tofNestart}  
\end{figure}

\section{Experimental procedure}

\subsection{Detector setup}
The experiments were carried out with the time-of-flight detector {\it COSY-TOF} located at an external beam line of 
the COoler SYnchrotron COSY (For\-schungs\-zen\-trum J\"ulich).
The COSY machine provides proton beams of very high quality (spill length $\approx$ 5 min; several $10^6$
protons/s; low emittance of $< 5\, \pi\, \mathrm{mm\, mrad}$; 
relative momentum uncertainty  $\Delta p/p < 10^{-3}$).

The layout of the {\it COSY-TOF} detector is shown in the upper part of fig.~\ref{fig:tofNestart}; 
in the lower part the near target region with the 
time-of-flight start and tracking detectors \cite{tofbeschreibung,Erlangenstart} is sketched.
The interaction volume is small and well defined as the narrow beam with 
Gaussian profile ($\sigma_{x,y}<300\,\mathrm{\mu m}$) is directed onto a liquid hydrogen target of only 4 mm length \cite{target}.
The emerging particles traverse just behind the target ($\approx 25$ mm) a 24-fold segmented 
scintillation detector (``start-detector'')
which provides the start signal for the time-of-flight measurement. 
At a distance of 30 mm downstream of the target a double-sided silicon-microstrip detector is installed, followed by
two double-layered scintillating fiber hodoscopes at 100 and 200 mm.  
These three tracking detectors measure the coordinates of traversing 
charged particles in three dimensions with a spatial resolution of $\approx 100\,\mu m$ 
(microstrip) and $\approx1.5$ mm (hodoscopes).
 
After a flight path of $\approx\negthinspace$ 3 m through the evacuated ves\-sel (0.2 Pa) all charged particles 
are detected in the highly segmented stop components. They consist of two triple-lay\-ered 
forward hodoscopes (central and ring hodoscope) \cite{forewardhodo}
and the barrel hodo\-scope \cite{barrel}, all manufactured from BC412 scintillating material. 
From the combined measurement of time and position the velocity vectors of 
charged particles (originating from the target) 
are determined with a time-of-flight resolution of better than $\sigma_{TOF}$ = 300 ps 
and an angular track-resolution of better than $\sigma_{\sphericalangle}=0.3^\circ$. 
Vertices from neutral particles decaying 
behind the microstrip detector and before the first hodoscope are reconstructed 
from the tracks of their two charged daughter particles 
with an accuracy of $\sigma_{x,y} < 1\,\mathrm{mm}$ and $\sigma_{z} < 3.0\,\mathrm{mm}$.
For the present analysis this feature is of crucial importance as a measurement
of the  four-momentum of the $K^0_S$ is mandatory for the full kinematic reconstruction of the final state.

The {\it COSY-TOF} detector stands out for its low mass areal density of target, start-detector, and tracking detectors. 
This renders
the influence of small angle scattering and energy loss almost negligible.
In addition, the {\it COSY-TOF} detector has a high efficiency of $>95\% 
$ for the detection of charged particles and covers a large solid angle
($1^\circ<\theta<60^\circ,\, 0^\circ<\phi<360^\circ$) in the laboratory frame.
These features allow the study of different reaction channels 
({\it e.g.} ~$\mathrm{pp\to pp}$ \cite{WU07}, $\mathrm{pp\gamma}$ \cite{huebner98}, $\mathrm{pp\eta}$ \cite{roderburg03}, 
$\mathrm{d\pi^+}$ \cite{MSWdiss}, $\mathrm{pp\omega}$ \cite{wu09}, 
$\mathrm{pK^0\Sigma^+}$ \cite{tof07}, and  $\mathrm{pK^+\Lambda}$ as well as $\mathrm{pK^+\Sigma^0}$ \cite{ddlambdasigma10})
from the same data sample by examining the measured time-of-flight of the charged particles and the overall event topology.

\subsection{Principle of measurement and data analysis}

The TOF start-detector setup was designed to provide an effective means for the analysis of 
final states with open strangeness ({\it e.g.} ~$pK^+\Lambda$, $pK^+\Sigma^0$, $pK^0\Sigma^+$). 
Here, two charged particles are emitted at the three particle production vertex.
In the case of $pp\to pK^0\Sigma^+$ the $K^0_L$ escapes the detector ($\mathrm{c}\tau_{K^0_L}$ = $15.3\,\mathrm{m}$) while the 
$K^0_S$ has a considerable probability 
($\mathrm{c}\tau_{K^0_S}$ = $26.8\,\mathrm{mm}$) to decay into two charged particles 
behind the start-detector and before the first hodoscope (see fig.~\ref{fig:tofNestart}, lower part).
Hence, the start-detector will
be hit by only two, while the detectors located further downstream, {\it i.e.} ~the two hodoscopes and the stop
detectors, will be hit by four charged particles. This ``multiplicity jump of charged particles'' is set as a 
trigger condition during data taking and is also the first condition required in the off-line analysis.

In order to discriminate the reaction $pp\to pK^0\Sigma^+$ from background, 
the characteristics of its final states is exploited: 1) a prompt track
emerging from the target (proton), 2) a decay ''V''($K^0_s\to\pi^+\pi^-$) with its vertex located 
in the decay volume (displaced vertex), and 3) one
additional hit somewhere in the stop detector (due to the  charged 
decay particle from either $\Sigma^+\to p\pi^0$ or $\Sigma^+\to n\pi^+$).

Primary track and decay V candidates are selected by applying the following conditions: 
a primary track must have a signal in the start-detector and in one of the stop components.
In addition, at least two signals from the three sub-detectors (microstrip detector, small and large hodoscope) are required. 
A secondary V must consist of two independent arms, each defined by fitting points  
in both hodoscopes and the stop detector. 
The point of closest approach of both arms
is considered to be the kaon decay vertex; the vector connecting the center of the target and this decay vertex is
considered to be the kaon's direction of flight ($\hat{p}_{K^{0}_{S}}$).
In order to discriminate a secondary V candidate against 
the background generated by primary particles inducing reactions in the start-detector a minimum angle 
of the pion with respect to
the mother particle of $10^\circ$ and a minimum angle between both pions of $30^\circ$
is required. 

All permutations of primary tracks and decay V candidates
are subjected to an overall quality check which includes the number of involved fitting points,
the distance of closest approach of both tracks of the decay V, the quality of all fittings procedures 
($\chi^2$ values), and the coplanarity of the decay V with respect to the primary kaon  
($\hat{p}_{K^0_S}\times(\hat{p}_{\pi^+}\times \hat{p}_{\pi^-}) \approx 0$).
The combination of a primary track and a secondary V with the best overall quality is kept for further analyses.
This method was developed by means of Monte Carlo data
which show that in the final event sample 86\%
of the events are reconstructed correctly.

So far only geometric information is exploited in order to identify the primary proton 
and the two decay pions of the primary (neutral) kaon. 
The velocity vector of a primary track is calculated from ($t^{stop} - t^{start}$) and by 
assigning the proton mass to this velocity vector the proton four-momentum is obtained. 

As the secondary pions from the $K^0_S$ decay emerge behind the start detector they lack an individual
start-time information. 
Nevertheless, their momenta can be calculated
from the kinematic relations between the measured angels of the daughter pions with respect
to their mother kaon. The sum of both pion four-momenta then yields the four-mo\-men\-tum of the kaon.
The four momenta of beam, target, kaon, and proton are used in order to calculate the
four momentum of the $\Sigma^+$ ($\mathcal{P}_{\Sigma^+} = \mathcal{P}_b+\mathcal{P}_t-\mathcal{P}_{K^0}-\mathcal{P}_p$). 
The invariant mass of the $\Sigma^+$ is the ``missing mass'' of the final state.
The resulting missing mass spectrum is the backbone of the analysis.

Although the two pions do not provide an individual start-time signal their stop-time information
can nevertheless be used as a means to substantially reduce experimental background \cite{LKdiss}.
For this purpose the difference of the time-of-flights of the two pions is calculated via:

\begin{eqnarray}
	\Delta_{tof}   & = & (t^{stop}_{\pi_1}-t^{start}_{\pi_1})-(t^{stop}_{\pi_2}-t^{start}_{\pi_2}) \\\nonumber
		                         & = &  t^{stop}_{\pi_1}-t^{stop}_{\pi_2},
  \end{eqnarray}
as both pions are created simultaneously ($t^{start}_{\pi_1}=t^{start}_{\pi_2}$). 
This measured time difference is then compared to the time difference determined from the 
path length of the pions in the detector and the pions' calculated momenta. 
Measured and calculated time differences must match with\-in $2\,ns$ ($|\Delta_{tof}^{calc}-\Delta_{tof}^{meas}| \le$ 2 ns).
This requirement does not change the number of identified events significantly in the final event sample. 
The amount of experimental background, however, is reduced by 
about a factor of two.

 It should be noted that the information of the $\Sigma^+$ enters 
only by the presence of a fourth charged particle {\it somewhere} in the detector.
The characteristic ``kink-angle'' (cf.~lower part of fig.~\ref{fig:tofNestart})
could be used in order to discriminate between the 
two main decay channels of the hyperon 
$\Sigma^+\to p\pi^0$ and $\Sigma^+\to n\pi^+$. 
In fact, due to the mass difference of the charged daughter particles 
90\% of the $p\pi^0$ ($n\pi^+$) decay branch is found with a kink-angle below (above) $12^\circ$ .
The ratio of the acceptance corrected counts is 
$N_{\Sigma^+\to p\pi^0}/N_{\Sigma^+\to n\pi^+}= 0.99 \pm 0.16$, 
{\it i.e.} ~fully compatible with the accepted ratio of the branching ratios 
($\mathcal{B}_{\Sigma^+\to p\pi^0}/\mathcal{B}_{\Sigma^+\to n\pi^+} = 1.07$ \cite{pdg06}).
However, no selection based on the kink-angle has been applied in the final analysis since it 
neither improves the missing-mass resolution nor increases 
the signal-to-background ratio substantially.

The effects of various background reactions on the final missing mass spectrum were studied by 
Monte Carlo simulations. 
For this purpose $10^7$ events, distributed according to equal population of the available phase-space,
were generated for each of the reactions 
$pp\to pp$, $d\pi^+$, $pp\pi^0$, $pn\pi^+$, $pp\omega$, $pK^+\Sigma^0$ and $pK^+\Lambda$
and subjected to the  $pp \to pK^0\Sigma^+$ analysis routines. All but the hyperon 
channels produce negligible ($<1/10^{6}$) or no background at all. 
The reactions $pp\to pK^+\Sigma^0$ and $pp\to pK^+\Lambda$
cause a broad missing mass spectrum; no peak at the $\Sigma^+$ mass is observed in either cases. 
The acceptance for both channels is about 1.5\% with respect to the $pK^0\Sigma^+$ channel.
Considering the total cross sections for the $\Lambda$ and $\Sigma^0$ channels 
\cite{ddlambdasigma10} their added contribution to the final event sample is 
below 15\% for each of the three measured beam momenta.

\subsection{Acceptance correction and absolute normalization}
The Monte-Carlo package used \cite{lasvegasbeschreibungBrand,lasvegasbeschreibungZielinsky} 
models the detector and the physical processes to great detail. 
The event generator produces the particles of the exit channel either according to 
three-body phase space,
or alternatively, intermediate resonances can be chosen in order to model a two-step production process 
($\mathrm{pp\to p}N^*, N^*\to\mathrm{K^0\Sigma^+}$).
The particles (and their daughters, granddaughters, etc.) are then propagated through the detector. 
Branching ratios and lifetimes of all particles are incorporated according to the values given in 
\cite{pdg06}. 
Energy loss, small-angle scattering, nuclear reactions, and $\delta$-electrons are considered. 
Digitized QCD- and TDC-signals are generated from the energy deposited in the active detector components.
Noise and thresholds are modelled as known from the measured detector response. 
Deviations from an homoge\-neous\-ly populated phase space can be introduced by weight functions 
on an event-by-event basis (this procedure is called {\it filtering} in the following). 
Finally, the Monte Carlo data are subjected to the same routines as real data in order 
to determine the acceptance\footnote{Throughout this paper, the term ``acceptance'' is used for the 
convolution of solid angle coverage, detector-, and reconstruction-efficiency.}. 

\begin{figure*}[ttt]
\resizebox{1.0\textwidth}{!}{ \includegraphics{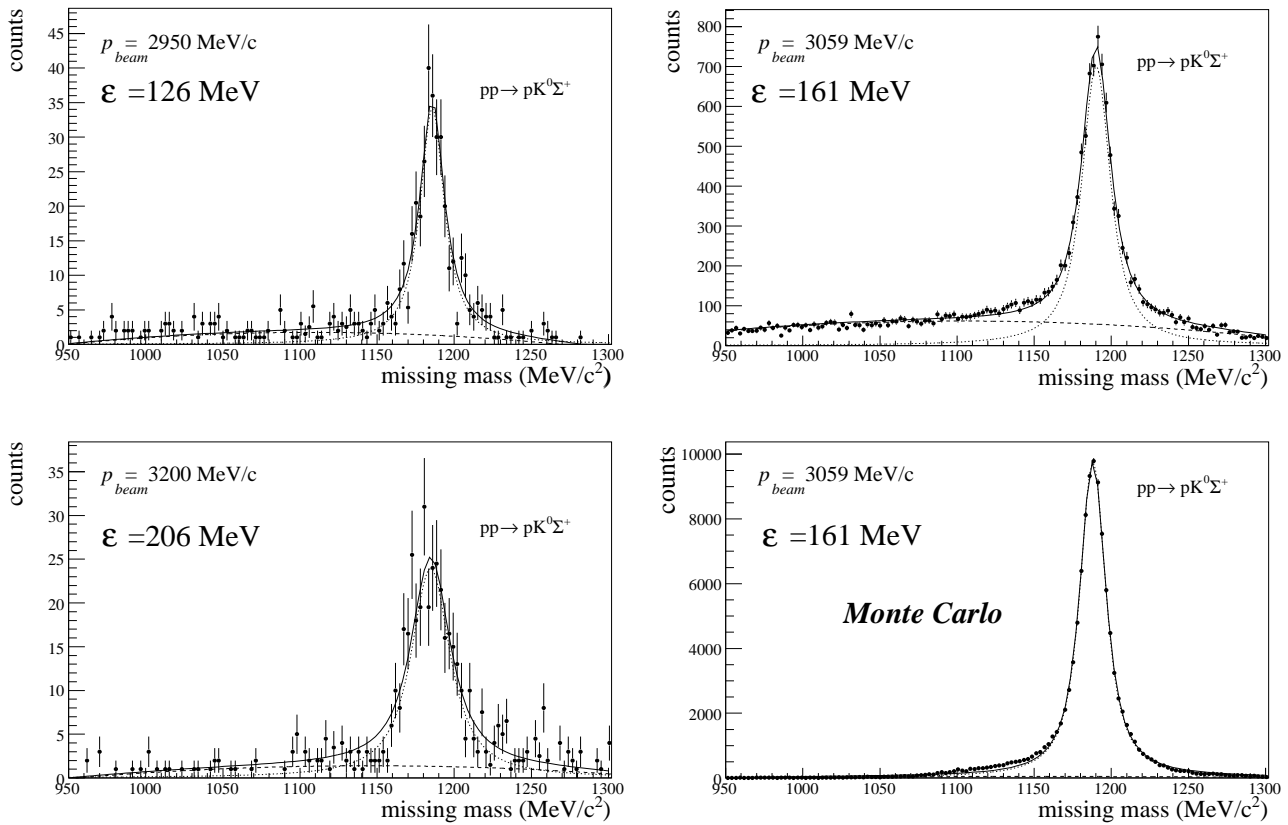} }
\caption{Proton-$K^0$ missing mass spectra measured for the three excess energies. 
A prominent peak is observed above a smooth background in all cases. 
The results of a fitting procedure for signal (dotted line), background (dashed line),
and total spectrum (solid line) are shown in all cases. In the lower right frame the Monte Carlo result 
for $\epsilon = 161$ MeV is depicted. }
\label{fig:viermalmm}  
\end{figure*}

The overall acceptance of about one percent for the reaction under study is determined
by the ratio of positively identified events to the number 
of generated ($10^7$) Monte Carlo events. This value is well explained by the event kinematics, 
the detector geometry, and the detector performance: 
the probability to observe the neutral kaon via its $K^0_{S}$ component (50\%),
the branching ratio of the kaon to two charged pions (69.2\%), the 
probability of the secondary vertex to be located within the 
fiducial ``decay volume'' (31\%), the probability of both secondary pions to traverse 
the active detector range (23\%), 
the reconstruction efficiency for both pions (60\%),
the reconstruction efficiency of 
a primary proton track (95\%), and the detection probablilty of the decay 
particle of the hyperon (82\%). The acceptance has been subject 
to a detailed reevaluation, resulting in a modification of the total cross section 
at $\epsilon$=161 MeV presented below compared to the previously published value \cite{tof07}.

The relative uncertainty of the overall acceptance correction was determined 
by considering the following effects:
1) The efficiency for the detection of charged particles is generally high
($>$95\%) and known with an uncertainty of  $\pm 5\%$ for each of the $\approx$ 1500
individual detector channels. This leads to a contribution to the uncertainty of the 
acceptance correction of 14\% for the present reaction. 
2) By altering all restrictions imposed during the data analysis the number of 
$pK^0_s\Sigma^+$ events in the final event sample can be changed by +10\%/-70\%
while the total cross section changes by less than 10\%. This value is taken 
as the contribution of the choice of the restrictions used in the analysis 
to the overall uncertainty of the acceptance correction.
3) The influence of the choice of intermediate nucleon resonances (MC input) 
and angular distributions (filter) was found to be small (2\% each), if 
the mass and width of the resonance and the asymmetry of the filter functions
are altered within the limits imposed by the measurements (see below).
Adding these contributions quadratically an overall systematic uncertainty of 18\% is obtained.

With regard to differential distributions the acceptance varies in all cases smoothly 
with the observable 
under consideration (details will be shown later when presenting the final results). 
Here, an additional uncertainty $\Delta a_i$ comes about for each bin $i$ due to the 
gradient of the acceptance function $a_i$. This is accounted for by choosing as uncertainty the 
mean change of $a_i$ with regard to its adjacent bins, {\it i.e.} ~$\Delta a_i = \frac{1}{2}(|a_i-a_{i-1}|/2+|a_i-a_{i+1}|/2)$.

The absolute normalization is determined via the ana\-lysis of elastic scattering, 
which was recorded simultaneously during the experiment.
Our results are normalized to  the elastic scattering cross sections from the 
EDDA collaboration \cite{edda00} and yielded time-integrated luminosities 
of $16.9\,\mathrm{nb^{-1}}$ ($\epsilon=126$ MeV), $214\,\mathrm{nb^{-1}}$ (161 MeV), 
and $6.4\,\mathrm{nb^{-1}}$ (206 MeV).
The total uncertainty of this procedure (5\%) is due in equal parts to our analysis and the 
uncertainty of the literature data. For details see \cite{MSWdiss,WU07}. 

\subsection{Determination of total and differential cross sections}
\label{signalyield}

Figure \ref{fig:viermalmm} shows the $pK^0$ missing mass spectra 
obtained for the three excess energies of 127, 161, and 206 MeV. 
Prominent signals for the $\mathrm{\Sigma^+}$ hyperon can be seen at
its central mass of 1189 MeV/$c^2$ above a smooth and structureless background. 
As usual for time-of-flight detectors, the missing mass resolution (momentum resolution) 
is best for smaller velocities in the exit channel (smaller beam momenta in the entrance channel).
When comparing the spectra one clearly notes the higher integrated luminosity for $\epsilon=161$.
At $\epsilon=126$ and $206$ $\mathrm{MeV}$, however, a better beam quality was available, as 
manifested by the lower background contribution. In the lower right frame 
the Monte Carlo result is shown for $\epsilon=161$ MeV.

The number of events in the missing-mass peak are obtained by consecutively fitting 
first the  background and then the signals.
The background is parameterized by a qua\-dra\-tic polynomial, where only missing masses
below and above the $\Sigma^+$-peak are taken into account ($m<1100\,\mathrm{MeV}/c^2$  and $m>1250\,\mathrm{MeV}/c^2$). 
Then, the background parameters are fixed and the signal is described by a Voigt function 
(convolution of a Gauss- and a Lorentz-function).
Voigt functions are chosen since they model properly
the signal shape of a rather narrow peak accompanied by broader tails. The integral of the Voigt function
represents the total number of positively identified events.
The overall systematic uncertainty due to signal and background separation is determined by varying 
the fit-region for the background-fit below and above the $\Sigma^+$-peak and was found to be $\pm4\%$.

The total cross sections for all three excess energies are then 
obtained from the number of identified events ($N$), the integrated luminosity ($\mathcal{L}$),
and the acceptance ($a$) according to $\sigma = N/(\mathcal{L}\cdot a)$.
As statistical uncertainties the numerical uncertainties of the fitting procedure of the signal 
will be quoted. As systematic uncertainties the quadratic sum of the uncertainties 
of luminosity determination (5\%), acceptance correction (18\%), and signal-to-background separation (4\%)
will be given.

Differential cross sections are determined in analogy to the total yield,
only that the amount of signal and background is determined from individual missing mass spectra 
generated for each bin of the observable under study. Also the uncertainty of the signal-background
separation is determinded individually.
Data will be shown if the uncertainty in a specific bin (root-mean-square of the statistical uncertainty combined with
the differential uncertainty of the acceptance correction and the signal-background separation) 
is below 80\% of its cross section value.

The numerical values of all total and differential cross sections will be given below in tables. For the latter
case the coefficients of least square fitting with Legendre polynomials
\begin{equation}
{\mathrm d}\sigma/{\mathrm d}\Omega=\sum _{l=0}^{l_{max}} a_l\cdot P_l,\quad{l=0,1,2}. 
\end{equation}
will be given in order to judge asymmetries ($P_1$, representative for 
all $P_{odd}$) and anisotropies ($P_2$).
It should be noted in passing that the integrals 
of all angular distributions 
($\sigma_{tot} = \int \frac{d\sigma}{d\Omega}d\Omega = 4\pi\cdot a_0$)  
match the total cross section within uncertainty.

\section{Results and discussion}
For a reaction of type $a+b \to 1+2+3$ the reaction space given by kinematics is 12 dimensional 
(three 4-momentum vectors of the exit channel), if the 
entrance channel is fixed (masses, $\sqrt{s}$).
Knowing the masses of the exit channel reduces the dimensionality to 9 (three 3-momentum vectors).
Four energy-momentum conservation equations reduce the dimensionality to five.  
If no spin direction is preferred, the azimuthal dependence is trivial 
and four dimensions (variables) remain in order to uniquely describe the reaction kinematics.

\begin{figure}[tbh]
\resizebox{0.5\textwidth}{!}{ \includegraphics{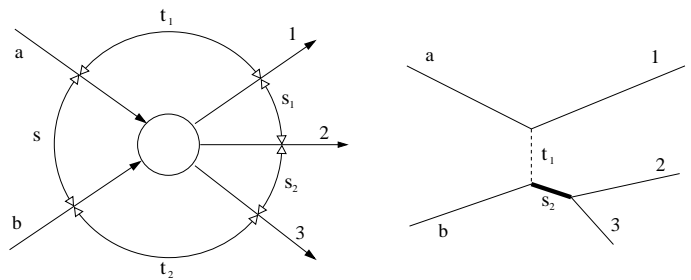} }
\caption{Graphical representation of $s$, $s_1$, $s_2$, $t_1$, $t_2$  following the prescription of 
ref.~\cite{Byckling} %(page 116-118) 
for the most general case of a $ab\to 123$ reaction (left), and for the case of
a two step reaction involving an intermediate resonance (right).
}
\label{fig:byckling}  
\end{figure}

The actual choice of the linearly independent basis of such a four dimensional space (or four experimental observables)
is not unique. One convenient choice are independent Mandelstam-like invariants \cite{Byckling}:
\begin{eqnarray}
  s_1 &\equiv& s_{12} = (\mathcal{P}_1+\mathcal{P}_2)^2 		  \\
  s_2 &\equiv& s_{23} = (\mathcal{P}_2+\mathcal{P}_3)^2  \nonumber    \\
  t_1 &\equiv& t_{a1} = (\mathcal{P}_a-\mathcal{P}_1)^2  \nonumber    \\
  t_2 &\equiv& t_{b3} = (\mathcal{P}_b-\mathcal{P}_3)^2  \nonumber ,
  \label{eqn:ttss}
\end{eqnarray}
where  the $\mathcal{P}_i$ denote the four-momentum of particle $i$. 
The physical meaning of these variables can be infered from fig.~\ref{fig:byckling}:
$t_1$, $t_2$ are the squared momentum transfers between the initial particles and two of the three ejectiles; 
$s_1$, $s_2$ are the squared invariant masses of two of the three final state subsystems. In case of resonant
production we adopt the convention that $t_1$ is the momentum carried by the exchange particle 
while $t_2$ is the momentum transfer from particle $b$ to one of the decay products of the resonance.
In the case of a symmetric entrance channel as in the current case particles $a$ and $b$ are interchangeable.

%-----------------------------------------------------------------

Using these four variables the four-fold differential
cross section can be written as

\begin{equation}
d^4\sigma/ds_1ds_2dt_1dt_2 = \Phi \cdot |M(s;s_1,s_2,t_1,t_2)|^2 ,
\label{eq:diffcross}
\end{equation}
where $\Phi$ represents the properly normalized phase-space factor and $M$ is the 
transition matrix element.

The four relativistic invariants in eq.~\ref{eqn:ttss} are linearly connected to specific angles, for instance,
$s_1$ is a linear function of cos($\angle(\vec{p_1},\vec{p_2})$) in a properly chosen reference frame.  
Similarly, other  angles can be used to substitute $s_2$, $t_1$, $t_2$.
Therefore, the four-fold differential cross section can also be written as a function of a combinations of angles and invariants.

The experimental goal is to provide the four-fold differential cross section (in whatever basis). This, however,
is unrealistic in many cases, simply for statistical reasons. 
In practice, one therefore needs to reduce the dimensionality by projecting onto subspaces:
Integration over $t_1$ and $t_2$ results in Dalitz plots. 
Projecting onto one dimension yields, for instance, 
invariant masses or angular distributions in particular reference frames.

The reduction of dimensionality 
is accompanied with a significant loss of information, as possible correlations can no longer be recognized. 
In addition, one has to be cautious about
the kinematic correlation of the three body final
state, as in specific cases a true physical cause in one variable can mimic a characteristic signal
in another variable \cite{berger69} (kinematic reflection).
It therefore is essential to evaluate as many as 
possible (linear independent) projections of the four-fold differential cross section. 
These projections are then to be described simultaneously by a theoretic model.

\subsection{Total cross sections}
\label{sigmatot}

Integration over all four variables in eq.~(\ref{eq:diffcross}) results in the total cross section.
Thus, a single value for the total cross section does not allow to infer 
any details of the reaction mechanism.
Nevertheless, the evolution of the total cross section 
with excess energy, the excitation function ($\sigma=\sigma(\sqrt{s})$),
is often used as a first means to judge how well (different) theoretical approaches are appropriate to 
describe these data.

\begin{figure}[ttt]
\resizebox{0.5\textwidth}{!}{ \includegraphics{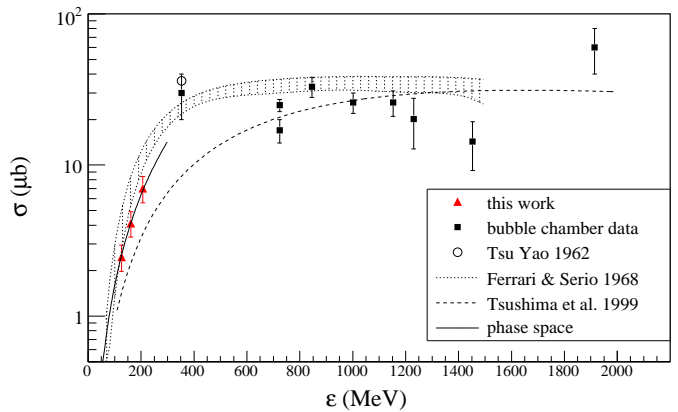} }

\caption{Total cross sections of the reaction $\mathrm{pp\to pK^0\Sigma^+}$. 
The triangles at low excess energies represent the present data. 
The solid square symbols are from bubble chamber experiments (compiled in \cite{baldini88}). 
The theoretical calculations shown are described in the text.}
\label{fig:worlddatatotalX}  
\end{figure}

The results for the total cross section are listed in ta\-ble \ref{tab:totalcrosssec}. 
They are included in fig.~\ref{fig:worlddatatotalX} which show the world data 
for the reaction $\mathrm{pp\to pK^0\Sigma^+}$. All data above 
$\epsilon = 300$ MeV stem from bubble chamber experiments \cite{Louttit61,bierman66,alexander67,Holmgren68,sondhi68,chinowsky68,Firebaugh68,Dunwoodie68,baldini88}. 
The data show a relatively large scatter, however indicate a rather constant total 
cross section over a wide energy range. The present 
data are the first to establish the excitation function at smaller excess energies. 

\begin{table}[ttt]
  \caption{Total cross sections for the reactions $\mathrm{pp\to pK^0\Sigma^+}$ for the three
  different excess energies. 
  The first uncertainty refers to statistical and the second to systematical ones.}
	\label{tab:totalcrosssec}
	  %\begin{tabular}{@{}llll}
	  \begin{tabular}{@{}cccc}
	  \hline\noalign{\smallskip}
	  $\varepsilon\;(\mathrm{MeV})$   &  acceptance (\%) &  signal (\#) & $\sigma_{\mathrm{tot}}\;(\mathrm{\mu b})$ \\
	  \noalign{\smallskip}\hline\noalign{\smallskip}
	  %$\mathrm{pp \to pK^{0}\Sigma}$ \\
	  $127$  & 0.93& 386  &  \makebox[0.5cm][r]{2.46}  $\pm$ 0.13 $\pm$ 0.47 \\ 
	  $161$  & 1.05& 9226 &  \makebox[0.5cm][r]{4.13}  $\pm$ 0.06 $\pm$ 0.79 \\
	  $206$  & 0.93& 412  &  \makebox[0.5cm][r]{7.02}  $\pm$ 0.36 $\pm$ 1.34 \\
	  \noalign{\smallskip}\hline
	\end{tabular}
\end{table}

The results of two early theoretical approaches are shown in the same figure.
In 1962 Tsu Yao \cite{tsuyao62} calculated the total cross section for the present reaction at $\epsilon=$ 353 MeV considering
solely single pion exchange. The resulting value of $\sigma = 36\,\mu b$ is in good agreement with the only experimental value
\cite{Louttit61} available at that time. 
The calculation by Ferrari and Serio in 1968 \cite{ferrari68} (based on the pioneering work of Ferrari \cite{ferrari60})
was aimed at the simultaneous description 
of various $pp\to NKY$ channels within a boson exchange
model (explicitly excluding nucleon resonances). 
The energy dependence of the total cross section calculated by this model is
shown as dotted lines for excess energies up to 1500 MeV. Here, the lower line represents the results for
purely pion exchange, while the additional contribution of kaon exchange (for a coupling constant of $G^2_{\Sigma}$
less than or equal to 1.6) is represented by the hatched area.
The calculation for pure pion exchange is in rather good agreement with our data.
Nevertheless, it will be shown in the following that nucleon resonances play an important role for the production 
of the $pK^0\Sigma^+$ final state. 
Hence, the model of Ferrari and Serio is lacking a central ingredient and the good agreement 
between the experimental data and theoretical prediction is probably a coincidence.

More recent calculations by Tsushima {\it et al.} \cite{tsuschima99} are based on a resonance model
where it is assumed that an excited baryon, $B^*$ ($N^*$ or $\Delta^*$),
is produced via meson ($\pi, \eta, \rho$) exchange ($pp\to pB^*$; $B^*\to K^+\Lambda$, $K^+\Sigma^0$, $K^0\Sigma^+$). 
With one set of parameters (coupling constants, cut-off values) all these three hyperon channels are treated on equal footing.
In the case of $pp\to pK^0\Sigma^+$ this model describes the total cross sections measured at excess energies 
above 800 MeV rather well (dashed curve), however misses
the experimental value at $\epsilon=$353 MeV by about a factor four.
This model underestimates the new data by a factor two.

%-----------------------------------------------------------------

We also include a solid line which represents the volume of available phase space
($\sigma\propto a\cdot \epsilon^2$, $a=1.617 \cdot 10^{-4}\,\mu b\, MeV^{-1}$). It describes
the data very well within the given experimental uncertainty. This indicates that 
the absolute value of the transition matrix element in eq.~\ref{eq:diffcross} does not depend strongly 
on excess energy in the investigated energy region. 
However, this does not imply that the transition matrix element itself is constant over the available phase space.

\subsection{Dalitz plot}
Integrating over $t_1$ and $t_2$ in eq.~\ref{eqn:ttss} results in the double-differential 
cross section $d^2\sigma/ds_1ds_2$, whose
re\-presen\-tation is the Dalitz plot \cite{dalitz53}. 
It connects two Lo\-ren\-tz invariants ({\it e.g.} ~$s_1=m^2_{12}$, $s_2=m^2_{23}$, squared invariant masses) and, 
hence, gives insight into the correlation of the three particle final state. 
If the  matrix element in eq.~\ref{eq:diffcross} is constant, {\it i.e.} ~does not depend on $s_1$, $s_2$, $t_1$, $t_2$,
the Dalitz plot is homogeneously populated. If, however, 
the matrix element is not constant, the deviation from an uniformly populated phase-space 
may show up as characteristic distortions. Prominent examples are  
resonances and FSI effects. If these signals are well separated, direct conclusions 
on physical properties can be drawn ({\it e.g.} ~mass and width of a resonance; scattering lengths via FSI).
This simple interpretation is not possible if different effects overlap on the Dalitz plot.
In this case a proper theoretical model (treating the dependence of the matrix element on all
four variables $s_1$, $s_2$, $t_1$, $t_2$)
must consider the different contributions coherently on the amplitude level.

%-----------------------------------------------------------------
\begin{figure}[ttb]
\resizebox{0.5\textwidth}{!}{ \includegraphics{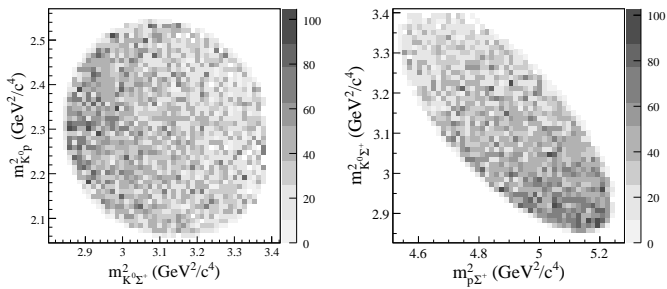} }
\caption{Dalitz plots of the squared invariant masses of the $K^0p$ vs $K^0\Sigma^+$ (left) and $K^0\Sigma^+$ 
vs $p\Sigma^+$ (right) 
subsystems for $\epsilon = 161\,\mathrm{MeV}$.
The data is acceptance corrected, however the background is not subtracted (see text).
In both frames a clear enhancement at low $K^0\Sigma^+$ masses indicate the presence of 
an intermediate resonance state ($B^*\to K^0\Sigma^+$) located near the lower mass boundary.
A contribution of a $p\Sigma^+$ FSI would lead to an enhancement 
of data at the lower mass boundary of the $p\Sigma^+$ mass axis. A significant signal of $p\Sigma^+$ FSI 
is not observed.
}
\label{fig:dalitz}  
\end{figure}

Acceptance corrected Dalitz plots ($m^2_{\mathrm{K^0p}}$ vs $m^2_{\mathrm{K^0\Sigma^+}}$ and $m^2_{\mathrm{K^0\Sigma^+}}$ 
vs $m^2_{\mathrm{p\Sigma^+}}$) are shown in fig.~\ref{fig:dalitz} for an excess energy of 
161 MeV. 
Data is shown for a missing mass region of $\pm 40\,\mathrm{MeV/c^2}$ around the $\mathrm{\Sigma^+}$ mass,
thereby reducing the contribution of background to below 20\%.
A bin-wise substraction of background is not feasible  for statistical reasons. 
However, sideband cuts show that the structures do not originate from the background.

In both Dalitz plots structures due to narrow resonances or final-state-interaction 
are not observed within resolution. However,  the relative bin occupancy of both Dalitz 
plots continuously increases towards low $K^0\Sigma^+$-masses.
This can be interpreted to be caused by one or more intermediate resonances 
($B^*\to K^0\Sigma^+$) with central masses ($m_i$) in the region of the 
lower mass boundary (or below) 
and broad widths ($\Gamma_i$) of at least $100\,\mathrm{MeV}$. 
The PDG \cite{pdg06} offers a whole list of
baryon resonances which could potentially interfere and contribute to the reaction. 
Due to the arguments given above, it is not trivial to extract the resonance 
parameters ($m_i, \Gamma_i$) from the Dalitz plots alone. 
In the following, however, we are going to argue that for the specific case of $pp\to pK^0\Sigma^+$ 
a set of one dimensional distributions is well suited to 
shed light on the reaction mechanism, particularly on the resonant contribution.

%-----------------------------------------------------------------
\begin{figure*}[ttt]
\resizebox{1.0\textwidth}{!}{ \includegraphics{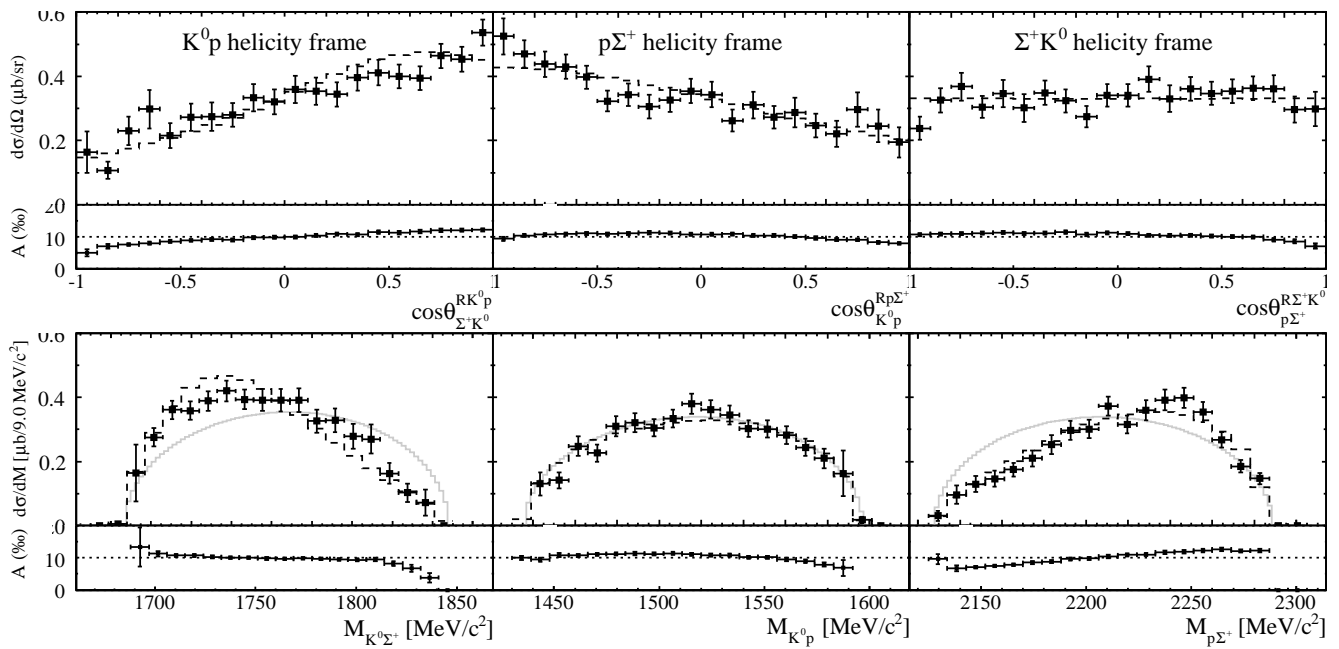} }

\caption{Angular distributions of the particles in helicity frames (top) and the invariant mass distributions
for all three two-body subsystems (bottom)
measured at an excess energy of $\epsilon=161$.
Error bars for each data point are the square root of the quadratic sum of the statistical, acceptance, and 
signal-to-background-separation uncertainty.
The dashed histograms represent the effect of an N(1720,150) intermediate resonance, which is 
used as input for the Monte Carlo simulation. 
The grey curves show the distibutions expected for a pure phase-space scenario. Below each distribution 
the differential acceptance is shown. 
The numerical values of the differential cross sections are listed in table \ref{tab:diffobsA} and \ref{tab:diffobsB}.
}
\label{fig:3059sigmaplusB}  
\end{figure*}

\subsection{Angular distributions in the helicity-frames and invariant mass distributions}
\label{invMNhelicity}
Invariant-mass distributions are obtained by projecting the Dalitz plot onto one of its axes, 
$s_1=m^2_{12}$, $s_2=m^2_{23}$, or $s_3=m^2_{31}$. 
Although the resonance parameters ($m$,$\Gamma$) are reflected in the invariant mass spectra,
their direct extraction is only possible for an isolated 
non interfering resonance.
In any case, one has to be cautious about kinematic reflections
as the three invariants are related via $s_1+s_2+s_3=s+m^2_1+m^2_2+m^2_3$.
Any structure deviating from phase space in {\it e.g.} ~$s_2$, will cause deviations of the pure phase-space behavior of
the other observables $s_1$ and $s_3$. It is therefore not a priori clear which 
invariant shows a cause and which an effect.
In contrast, the situation is more clear when analyzing the angular distributions in  helicity frames.

Angular distributions in helicity frames\footnote{
For reactions of type $ab\to \it 123$ 
the {\it 23}-helicity-frame ($R23$, $R$ indicates $R$eference frame) is defined as the
rest frame of the particles {\it (2,3)} ($\vec{p}_3=-\vec{p}_2$).
The respective (polar) helicity angle in this frame is defined as the angle between particle 
$3$ and $1$ ($\theta^{R23}_{13}$). By cyclic permutation three helicity frames can be constructed  for the 
three-body final state ({\it {R23, R31, R12}}).
}
are essentially special projections of a Dalitz plot.
There is a linear relation between  $s_i$ and 
$\mathrm{cos}\theta^{Rjk}_{ij}$ (and cyclic permutations), therefore $s_1$ and $s_2$ 
can be substituted by two helicity angles in eq.~\ref{eqn:ttss} \cite{Byckling}.

For a single resonance (in the following we assume it to decay into the $23$-system) 
the properties of mass and width
lead to anisotropic distributions in the 12- and 13-helicity frames,
while the angular distribution in the 23-helicity frame remains
isotropic (projection of the ``resonance band'' in the Dalitz plot).
This isotropy in the 23-helicity frame is independent of mass and width
of the resonance and therefore uniquely identifies the decay channel.
This statement is also valid if more than one resonance (decaying into the same two-body system)
contribute incoherently.

The angular distributions in all three helicity frames 
measured at an excess energy of $\epsilon=161\,\mathrm{MeV}$ are shown in the upper row of fig.~\ref{fig:3059sigmaplusB}.
In the lower row the spectra of all three invariant-mass subsystems are displayed.  
The acceptances shown under each distribution are rather constant 
for all spectra. Due to the  the method 
of determining the yield the physical background is subtracted individually 
for each data bin (cf.~sec.~\ref{signalyield}).

It is obvious that the distributions in the $p\Sigma^+$ and $K^0p$ helicity 
frames are asymmetric (i.e.~$a_1 \neq 0$) while that in the $\Sigma^+K^0$ frame is almost isotropic 
($a_1 \approx 0$). This can be infered quantitatively from the values of the Legendre 
polynomials coefficients
listed in table \ref{tab:fitparasS}. 
It is also evident that the $K^0\Sigma^+$ invariant mass distribution deviates from that given 
by phase space (grey curve in fig.~7) and shows a clear  enhancement towards smaller masses; thus, 
strongly advocating a production scenario involving
a broad intermediate resonance with a central mass near (or below) the lower $K^0\Sigma^+$ mass boundary.

The isotropic distribution observed in the $K^0\Sigma^+$ helicity frame is of particular relevance
in case of resonant production: 
Firstly it indicates that the decay channel of the resonance is $B^* \to K^0\Sigma^+$.
Secondly, there is no preferred orientation of the $B^*$-spin with respect to the 
$B^*$ direction of flight. This is an indication of the presence of several partial waves.
And thirdly, the observed isotropy signifies that {\it only one} resonance participates or, 
if more than one resonance is involved, 
they {\it do not interfere} (except for the possibility that the interference pattern mimics an 
isotropic distribution). 

In the following, we assume that only one resonance participates.
Indeed, all six distributions of fig.~\ref{fig:3059sigmaplusB}  are well described by
the results of a Monte Carlo simulation (dashed histograms), where 
the three-particle final-state is modeled via 
$\mathrm{pp\to pN^*, N^* \to K^0\Sigma^+}$.
The central mass and the Breit-Wigner width of the resonance were chosen to be $m_{N^*}=1720\,\mathrm{MeV/c^2}$ and
$\Gamma = 150\,\mathrm{MeV/c^2}$ (abbreviated in the following as 
N(1720,150)). 

\begin{table}
\caption{Legendre polynomial coefficients (in units of $n$b/sr) 
determined by least square fitting to 
angular distributions of the reaction $pp\to pK^0\Sigma^+$ 
at $\epsilon=161\,\mathrm{MeV}$, in the overall CMS, the Jackson and helicity frames (top to bottom).
}		\begin{tabular}{@{}lrrr}
		\hline\noalign{\smallskip}
  cos & $a_0\;\;\;\;\;$ & $a_1\;\;\;\;\;$ & $a_2\;\;\;\;\;$ 	\\  		\noalign{\smallskip}\hline\noalign{\smallskip}
   $\theta^*_p$ 						  &  $  335 \pm \makebox[0.32cm][r]{ 20} $ &   $	3 \pm \makebox[0.32cm][r]{ 37} $ &   $   59 \pm \makebox[0.32cm][r]{ 39} $   \\ \noalign{\smallskip}
   $\theta^*_K$ 						  &  $  330 \pm \makebox[0.32cm][r]{ 22} $ &   $ -10 \pm \makebox[0.32cm][r]{ 44} $ &	$  109 \pm \makebox[0.32cm][r]{ 39} $	\\ \noalign{\smallskip}
   $\theta^*_{\Sigma^+}$				  &  $  325 \pm \makebox[0.32cm][r]{ 11} $ &   $   22 \pm \makebox[0.32cm][r]{ 20} $ &   $   46 \pm \makebox[0.32cm][r]{ 29} $   \\ \noalign{\smallskip}\hline\noalign{\smallskip}
   
   $\theta^{RKp}_{bK}$  				  &  $  319 \pm \makebox[0.32cm][r]{  6} $ &   $   38 \pm \makebox[0.32cm][r]{ 12} $ &   $   52 \pm \makebox[0.32cm][r]{ 16} $   \\ \noalign{\smallskip}
   $\theta^{Rp{\Sigma^+}}_{bp}$ 		  &  $  322 \pm \makebox[0.32cm][r]{  7} $ &   $ -27 \pm \makebox[0.32cm][r]{ 14} $ &	$	76 \pm \makebox[0.32cm][r]{ 20} $	\\ \noalign{\smallskip}
   $\theta^{R{\Sigma^+}K}_{b{\Sigma^+}}$  &  $  322 \pm \makebox[0.32cm][r]{  7} $ &   $  -7 \pm \makebox[0.32cm][r]{ 13} $ &	$	99 \pm \makebox[0.32cm][r]{ 16} $	\\ \noalign{\smallskip}\hline\noalign{\smallskip}
   
   $\theta^{RKp}_{{\Sigma^+}K}$ 		  &  $  323 \pm \makebox[0.32cm][r]{  9} $ &   $  183 \pm \makebox[0.32cm][r]{ 16} $ &   $ -20 \pm \makebox[0.32cm][r]{ 20} $	\\ \noalign{\smallskip}
   $\theta^{Rp{\Sigma^+}}_{Kp}$ 		  &  $  326 \pm \makebox[0.32cm][r]{  9} $ &   $ -130 \pm \makebox[0.32cm][r]{ 17} $ &   $   41 \pm \makebox[0.32cm][r]{ 22} $   \\ \noalign{\smallskip}
   $\theta^{R{\Sigma^+}K}_{p{\Sigma^+}}$  &  $  332 \pm \makebox[0.32cm][r]{  9} $ &   $   19 \pm \makebox[0.32cm][r]{ 17} $ &   $ -19 \pm \makebox[0.32cm][r]{ 22} $	\\ \noalign{\smallskip} 
	\noalign{\smallskip}\hline
	\label{tab:fitparasS}	
	\end{tabular}
\end{table}

The choice of this particular N(1720,150) resulted from a series of Monte Carlo simulations 
which were performed to study the kinematical effect of mass and width of different resonances such as 
N(1400,270) \cite{luizou06}, N(1650,300), N(1720,150), and N(1900,300)
chosen from \cite{pdg06}. We also included an N(1800,200) state in order to study the 
development of the kinematical effect with a narrower spacing of masses.
(As kinemantics  does not depend on isospin, the following statements apply also
for  $\Delta(1600)$, $\Delta(1620)$, $\Delta(1700)$,
$\Delta(1750)$, and $\Delta(1900)$ \cite{pdg06}.)

\begin{figure}[hbt]
\resizebox{0.5\textwidth}{!}{ \includegraphics{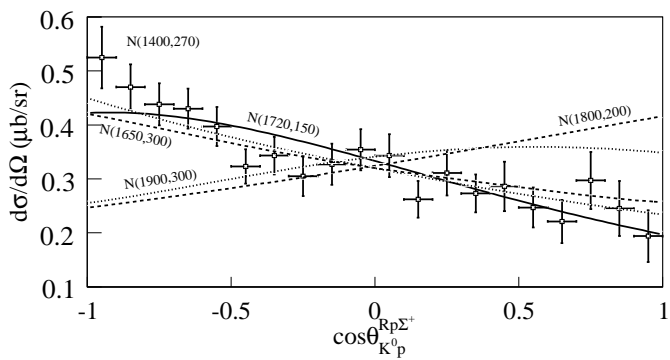} }
\caption{Angular distribution measured in the $p\Sigma^+$ helicity frame.
The lines are the Monte Carlo result for various resonances ($N(m,\Gamma)$). 
Simulations with low mass resonances reproduce the measured data with similar
quality, whereas heavy resonances clearly fail. The resonance with the best overall 
agreement between data and Monte Carlo is the N(1720,150) 
which is plotted as solid line.}
\label{fig:finalbild2}  
\end{figure}

As an pictorial example of such an analysis 
the distributions obtained in the $p\Sigma^+$ helicity helicity are shown in fig.~\ref{fig:finalbild2}.
Due to kinematics, all heavy resonances have a positive slope in the $p\Sigma^+$
helicity frame. They clearly fail to describe the data.
In contrast, all distributions resulting from the decay of light resonances display negative slopes in this 
helicity frame
and describe the data with similar accuracy.

In order to assess the quality of
the overall description of the data by each of these $N^{*}$ resonances Monte Carlo simulations
a simple quality criterion 
- root-mean-square of the error-weighted difference of data and Monte Carlo results, averaged
over all bins for the six observables of fig.~\ref{fig:3059sigmaplusB} - was applied.
It turns out that this figure of merit varies by about 10\% for the three light
resonances, where the N(1720,150) yields the best averaged description. 
This resonance is used throughout the whole analysis as Monte Carlo input for acceptance correction.

The working hypothesis of a single resonance governing the $pK^0\Sigma^+$ reaction process is further applied
in the next subsection.

%-----------------------------------------------------------------

%-----------------------------------------------------------------
\begin{figure*}[ttt]
\resizebox{1.0\textwidth}{!}{ \includegraphics{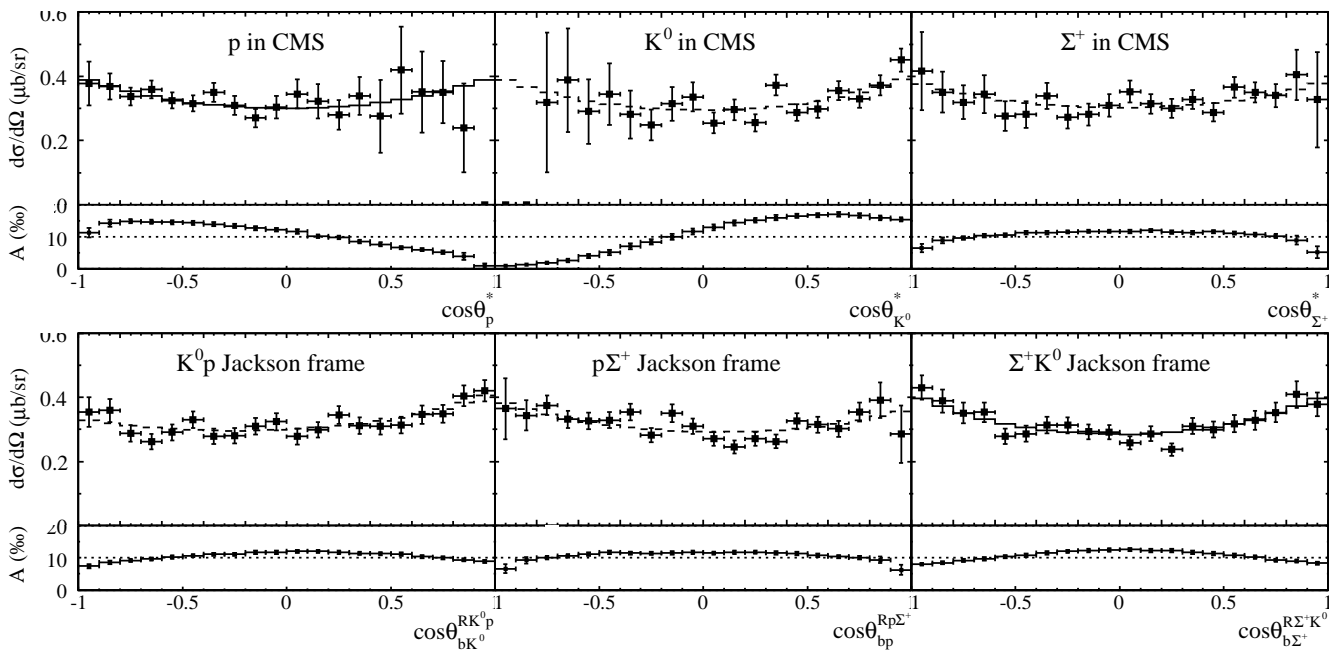} }
\caption{Angular distributions of the particles in the overall CMS (top) and the Jackson frames (bottom),
 for the reaction $\mathrm{pp\to pK^{0}\Sigma^+}$ 
measured at an excess energy of $\epsilon=161\,\mathrm{MeV}$ ($p_{beam}=3059\,\mathrm{MeV/c}$).
Error bars for each data point are the square root of the quadratic sum of the statistical, acceptance, and 
signal-to-background-separation uncertainty.
The solid histogram in the proton CMS and $\mathrm{\Sigma^+K^0}$ Jackson frame represents the respective 
Legendre polynomials of table \ref{tab:fitparasS} which are used as MC filter. Their affect on all other angular distributions 
are shown
by the dashed histograms. Below each angular distributions the differential acceptance is shown. 
The numerical values of the differential cross sections are listed in table \ref{tab:diffobsB}.
}
\label{fig:3059sigmaplusA}  
\end{figure*}

\subsection{Angular distributions in the CM- and Jackson-frames}
Angular distributions in the overall CMS can be described by Legendre polynomials.
In the specific picture of a production mechanism involving an intermediate resonance 
($pp\to pB^*$) the CMS scattering angle of the proton directly reflects the angular momenta involved in
the $pB^*$ channel. There is a linear dependence of the cosine of the scattering
angle on momentum transfer $t_1$, thus the proton CMS angle
can be taken as one of the linear independent variables in eq.~\ref{eq:diffcross}.
Due to the symmetric pp entrance channel all CMS distributions must of course be symme\-tric with 
respect to $\mathrm{cos}\theta^*=0$.

Jackson frames\footnote{\label{GJerklaerung} For reactions of type $ab\to \it 123$ the Jackson frame is defined as the
Lorentz frame in which the center of mass of the particles {\it (2,3)} is at rest ($\vec{p}_3=-\vec{p}_2$). 
The 23-Jackson frame and the 23-helicity frame are in fact the same Lorentz frame ($R23$). The word ``frame''
rather refers to the choice of the reference axis which in case of Jackson frames is 
the direction of the beam particle.
The Jackson angle is then defined as the angle between the beam direction and that of particle {\it 3} ($\theta^{R23}_{b3}$).
By cyclic permutation three Jackson frames can be constructed  for the 
three-body final state ({\it {R23, R31, R12}}).
}
are the natural Lorentz invariant frames to investigate the 
angular momenta involved in a two-particle subsystem \cite{GJ}.
To illustrate this, consider the secondary decay ($B^*\to K^0\Sigma^+$) at the $\mathrm{\pi p\to K^0}\Sigma^+$ vertex in 
fig.~\ref{fig:reactionmechnism}b, c; this represents a ``$2 \to 2$'' reaction. 
In this picture, the inverse reaction ($\mathrm{K^0}\Sigma^+\mathrm{\to  p\pi}$) must have the same properties 
due to time reversal invariance.
If one now imagines colliding beams of kaons and hyperons with $\vec{p}_{Y} = -\vec{p}_{\mathrm{K}}$, 
which is by definition the $\mathrm{K}Y$ Jackson frame, it is self-evident that the distribution of angles 
$\theta^{RKY}_{bK}$ 
of the (in this case emerging) proton with respect to the 
(in this case beam-axis defining) kaon contains information on the relative angular momenta involved.
There is a linear relation between $t_2$ and $\mathrm{cos}\theta^{RK^0\Sigma^+}_{b\Sigma^+}$, {\it i.e.} ~the $K^0\Sigma^+$ Jackson 
angle can substitute the momentum transfer between the beam particle and one of the resonance 
decay particles in eq.~\ref{eq:diffcross}. 
It is important to notice that the Jackson frame is a different Lorentz frame than the CMS and
no symmetry with respect to $\mathrm{cos}\theta=0$ is required.
However, the distributions measured with respect to the target and beam proton must be identical.

The angular distributions of all primary particles in the CMS measured at an excess energy of 
$\epsilon=161\,\mathrm{MeV}$  
are shown in the upper row of fig.~\ref{fig:3059sigmaplusA}.
The lower row shows the angular distributions in all three Jackson frames.  
The variation of the acceptance is shown under each distribution. 
In case of the proton and kaon the acceptance shows a quite strong
angular dependence which  is due to the strongly decreasing probability of
the kaon to reach the decay volume when emitted further into the backward CMS hemisphere ($\mathrm{cos}\theta^*_{K^0}\to -1$).
This experimental effect is mirrored 
in the proton acceptance distribution, as protons and kaon tend to 
be emitted into opposite hemispheres. 
For the hyperon, constructed from the combined momentum vectors of kaon and
proton, a much weaker angular dependence of its acceptance is found. Nevertheless, after acceptance
correction all spectra show the necessary symmetry with respect to $\mathrm{cos}\theta^* = 0$ ($a_1=0$), 
as the values of $a_1$ given in table \ref{tab:fitparasS} are all compatible with zero within uncertainty.
The $a_2/a_0$ ratio changes by less than 7\% if $a_1$ is forced to be zero.

A clear anisotropy is observed for all Jackson-frame distributions, as indicated 
by the $a_1$ coefficients listed in table \ref{tab:fitparasS}.
There is no need to introduce Legendre polynomials of order higher than two.
In the picture of a two step production process, this finding for the proton CMS
angular distribution and the distribution in the $K^0\Sigma^+$ Jackson frame
indicates that angular momenta of $l\le 1$ are involved
at both the $pp\to pB^*$ and $B^*\to K^0\Sigma^+$ subprocess.

In order to corroborate the two step production scenario we now turn to the simultaneous description 
of the data by Monte Carlo simulations, the result of which is shown as either 
solid or dashed histograms in fig.~\ref{fig:3059sigmaplusA}.
The solid histograms shown represent Monte Carlo data
which were filtered by means of weight functions in the two natural frames described above: 
The first filter was tailored to reproduce the measured proton CMS distribution 
(solid line in fig.~\ref{fig:3059sigmaplusA}, upper left frame).
This filter significantly affects via kinematic correlation the two Jackson frames
containing the proton (distribution in the $pK^0$ and $p\Sigma^+$  Jackson 
frames, dashed line in fig.~\ref{fig:3059sigmaplusA},
lower left and lower middle frame),
while barely affecting the kaon and hyperon CMS distributions.
The angular distribution in the $K^0\Sigma^+$ Jackson frame is not affected at all.
Hence it is justified to apply a second filter which 
is tailored to reproduce the measured distribution in the $\Sigma^+K^0$ Jackson frame 
(solid line in fig.~\ref{fig:3059sigmaplusA}, lower right frame). 
This filter significantly affects via kinematic correlations the kaon and hyperon 
CMS distributions (dashed lines in fig.~\ref{fig:3059sigmaplusA}, upper-middle and upper-right frame)
while barely affecting the distribution in the $p\Sigma^+$ and $pK^0$ Jackson frames.
The proton angular distribution in the CMS is not affected at all. 
The observation that both filters do not effect each other 
is a consequence of the linear independence of $t_1$ and $t_2$.

%-----------------------------------------------------------------
\begin{figure}[hht]
\resizebox{0.5\textwidth}{!}{ \includegraphics{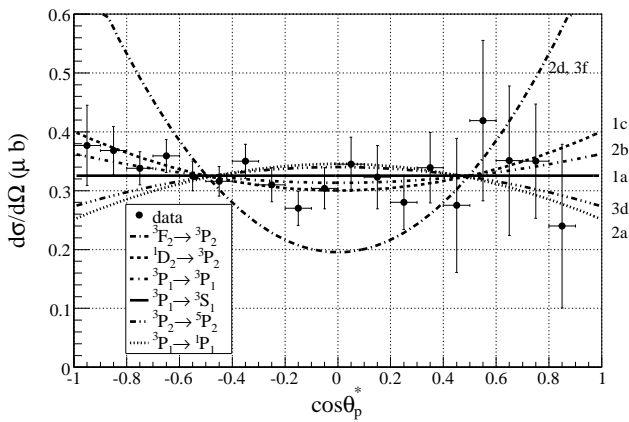} }

\caption{
Selected theoretical angular distributions calculated for the $pp\to pN^*$ reaction 
($N^*: S_{11}, P_{11}, P_{13}$) according to \cite{BlattBiedenharn}.
All distributions are of the type $d\sigma/d\Omega = a_0 +a_2P_2(cos\theta$), {\it i.e.} ~$l_{pN^*}\le 1$.
}
\label{fig:cmsblattBiedenharn}  
\end{figure}

Two other filter combinations also fulfill the requirement of linear 
independence ($K^0$-CMS and $p\Sigma^+$ Jackson frame;
$\Sigma^+$-CMS and $pK^0$-Jackson frame). Although these filter combinations represent somewhat 
awkward reaction scenarios ({\it i.e.} ~$pp\to K^0X$, $X\to p\Sigma^+$ and $pp\to \Sigma^+X$, $X\to pK^0$) they
are  not excluded a priori. Therefore both combinations were tested for the sake of 
completeness with the method introduced in sec.~\ref{invMNhelicity}.
Both filter combinations yield a description of the data 
slightly inferior to the filter combination discussed above 
(the ones shown as solid histograms in fig.~\ref{fig:3059sigmaplusA}).

\begin{table*}
\begin{center}
\caption{Possible transitions for $pp\to$ $pN(S_{11})$, $pN(P_{11})$, $pN(P_{13})$ 
which yield differential cross sections described by $d\sigma/d\Omega = a_0 +a_2 P_2(cos\theta)$ 
with all $a_2$ coefficients being $\neq 0$ but those of transition 1a and 3e which have $a_2=0$.
Note that initial partial waves of type $^1S_0$ and $^3P_0$ result in isotropic angular distributions
irrespectively of the intermediate resonance.
}

\begin{tabular}{@{}lcccccccc}
  		\hline\noalign{\smallskip}
  &                &          a                 &              b            &               c		       &     d	                   & e                         & f      & g\\ \noalign{\smallskip}
  		                                                                                                      	                        	 \hline\noalign{\smallskip}
1 &$pp \to pS_{11}$&  $^3P_1 \to  ^3S_1$& $^3P_1 \to  ^3D_1$&  $^1D_2 \to  ^3P_2$  &	                       &                           &                         &  \\ \noalign{\smallskip}
2 &$pp \to pP_{11}$&  $^3P_1 \to  ^1P_1$& $^3P_1 \to  ^3P_1$&  $^3P_2 \to  ^3P_2$  & $^3F_2 \to  ^3P_2$&                           &                         &  \\ \noalign{\smallskip}
3 &$pp \to pP_{13}$&  $^3P_1 \to  ^3P_1$& $^1P_1 \to  ^5P_1$&  $^3P_2 \to  ^3P_2$  & $^3P_2 \to  ^5P_2$& $^1D_2 \to  ^5S_2$& $^3F_2 \to  ^3P_2$& $^3F_2 \to  ^5P_2$ \\ \noalign{\smallskip}
   \noalign{\smallskip}\hline
	\label{tab:partialwaves}
\end{tabular}
\end{center}
\end{table*}

In the following it is assumed that both the distribution in the $K^0\Sigma^+$ Jackson frame 
and that of the proton in the overall CMS reflect the physical cause and the other four
distributions are simply their kinematical reflections. In addition the discussion will be based on the properties
of resonant $K^0\Sigma^+$ production from an initial state of identical particles.
The angular distribution in the Jackson frame reveals that the 
$K^0\Sigma^+$ subsystem carries angular momentum of $l\le 1$ (cf. table \ref{tab:fitparasS}).

It then follows from isospin, angular momentum, and parity conservation 
that only nucleon resonances of the type $S_{11}$, $P_{11}$, and $P_{13}$ 
and $\Delta$-resonances of the type $S_{31}$, $P_{31}$, and $P_{33}$ 
can be involved. All other resonances have decay angular momenta of $l>1$.

In the next step, all angular distributions for protons in the CMS are calculated 
for $pp\to pN(S_{11})$, $pN(P_{11})$, and $pN(P_{13})$ applying the formalism 
of Blatt and Biedenharn \cite{BlattBiedenharn}. Entrance channel partial waves  
up to $^3F_4$ are considered (spectroscopic notation $^{2S+1}L_J$, with 
$S$: channel spin, $L$: orbital angular momentum of the two-body system, 
$J$: total angular momentum of the channel). Angular distributions arising 
from $^1S_0$ and $^3P_0$ entrance channel partial waves are 
isotropic irrespective of the final state due to $J=0$ for the entrance channel.

All other transitions are listed in table \ref{tab:partialwaves}. Interestingly, 
all angular distributions resulting from these transitions are of the type 
$d\sigma/d\Omega = a_0+a_2P_2(\mathrm{cos}\theta_{CM})$, {\it i.e.}~no term 
higher than $P_2$ is involved which is perfectly in line with the experimental 
finding of $l\le 1$. The calculations according to ref.~\cite{BlattBiedenharn}  
yield $a_2=0$ for transition 1a and 3e (isotropy), $a_2 > 0$ (concave shape) 
for 1c, 2b, 2c, 2d, 3a, 3c, 3f and $a_2 < 0$ (convex shape) for 1b, 2a, 3b, 3d, 3g. 

In the case of $\Delta$-resonances isospin selectivity requires their production
from a spin singlet state (isospin triplet state) in the entrance channel, thus 
ruling out entrance channel $P$- or $F$-wave contributions. It is found that only 
two transitions are possible which yield angular distributions with $l \le 1$ 
namely $(^1D_2)_i \to (^3P_2)_f$ involving the $\Delta(S_{31})$ intermediate
state and $(^1D_2)_i \to (^5S_2)_f$ via the $\Delta(P_{33})$, the latter being 
isotropic. A $\Delta(P_{31})$ resonance yields angular distributions with an
$a_4P_4$ term indicative of $l=2$ which is not observed; 
hence such a resonance is excluded from the further discussion.

Examples of calculated distributions resulting from transitions listed in table \ref{tab:partialwaves} 
are shown in fig.~\ref{fig:cmsblattBiedenharn} normalized to the experimental data.
The concave curves 1c and 2b match the data best.
However, it is very unlikely that only one or two specific transitions govern the reaction process.
In fact, an inspection of table \ref{tab:partialwaves} reveals that 
the same initial and final state partial waves can occur, 
however with  different intermediate resonances
being involved ({\it e.g.} 2c - 3c, 2d - 3f). In addition, various initial state partial
waves pass through a particular resonance but end up in different final state partial waves ({\it e.g.} ~2a-2b, 3c-3d). 
A rather complicated scenario turns up and it is therefore impossible to pin down the 
individual contributions of particular transitions within this simple reaction model.
The question of which partial waves dominate the reaction can only be answered 
by a complete partial wave analysis.

Nevertheless, fig.~\ref{fig:cmsblattBiedenharn} indicates that a proper superposition of 
transitions will match the experiment data.
Thus, the working hypothesis introduced in Sect. \ref{invMNhelicity} 
of a reaction mechanism involving an intermediate resonance
is now limited to the states $S_{11}$, $P_{11}$, $P_{13}$, $S_{31}$, and $P_{33}$.

The clear non-zero angular momentum
observed in the $K^0\Sigma^+$ Jackson frame requires considerable strength of 
participating resonances with intrinsic angular momentum, {\it i.e.}~favoring 
$P_{11}$ and/or $P_{13}$ and/or $P_{33}$. Resonances with these spins and parities 
can be related to those listed by the PDG. It turns out that $N(1710)P_{11}$, 
$N(1720)P_{13}$, and $\Delta(1600)P_{33}$ are the only candidates with appropriate 
mass and a $K\Sigma$ decay branch. Exactly this type of resonance 
($m\approx 1720$ MeV, $\Gamma \approx 150$ MeV or 
$m\approx 1650$ MeV, $\Gamma \approx 300$ MeV) was also favored in describing 
the invariant mass and helicity angular distributions.
As no $\Delta(S_{31})$-resonance is known in this mass range which decays 
into $K\Sigma$, the $N(1650)S_{11}$ is the only further intermediate resonance to be 
possibly involved.

It must be emphazised that the results for mass and width of a potential intermediate resonance
obtained in sect.~\ref{invMNhelicity} and in this section resulted from exploiting 
orthogonal parameter spaces, namely $s_1$, $s_2$, $s_3$ and $t_1$, $t_2$, respectively.
The possibility of cross-checking results obtained in various reference frames and their combined interpretation 
underlines the importance of an analysis of the full four-dimensional reaction space.

\section{Summary}

The reaction $pp\to pK^0\Sigma^+$ was studied at the excess energies of $\epsilon =$ 126, 161 and 206 MeV
with the COSY-TOF spectrometer.  
The large acceptance allows the four-fold differential cross section 
$d^4\sigma/ds_1 ds_2 dt_1 dt_2$ to be determined. This high dimensionality however, needs to be reduced by
projecting onto a reduced number of observables to highlight certain aspects of this four-dimensional parameter space.
In doing so, Dalitz plots, invariant mass spectra, angular distributions in helicity frames, the overall CMS, and  
Jackson frames are obtained for $\epsilon =$ 161 MeV.

The measured total cross sections are the first to build the excitation function 
between the production threshold and the energy region covered by early bubble chamber experiments. 
The observed increase of the total cross section with excess energy is consistent with the increase 
of phase space volume. This, however, does not mean that the reaction is governed by phase space; it rather implies
that the transition matrix element does not (strongly) depend on energy up to about $\approx$ 200 MeV.  
The bubble chamber data show the total cross section to be rather constant ($\approx 20\,\mu b$)
in the energy range from 300 to 1500 MeV. 

The Dalitz plots do not show any sign of final-state-interactions. However, a clear signal for
resonant contributions ($pp\to pB^*$; $B^* = N^*, \Delta^*$; $B^*\to K^0\Sigma^+$) 
is found near the lower $K^0\Sigma^+$ mass boundary. This qualitative result is confirmed by the quantitative 
analysis of invariant mass spectra and angular distributions in helicity frames. The best description of the data by Monte Carlo is found for 
resonances with $m_{N^*} \approx 1720\,\mathrm{MeV}/c^2$ and $\Gamma\approx 150\,\mathrm{MeV}/c^2$. However, a contribution of 
lighter $N^*$ or $\Delta^*$ resonances cannot be excluded. In contrast,
the data clearly rule out a significant contribution of resonances with masses larger than 1800 $MeV/c^2$. 
The isotropic distribution in the $K^0\Sigma^+$ helicity frame shows again the absence of $p\Sigma^+$ 
final-state interactions as well as the absence of interference effects which would distort this distribution. 
The absence of interference allows the production mechanism to be interpreted
as a two step process $pp\to pB^*$, $B^*\to K^0\Sigma^+$ involving only one resonance.

The decay angular momentum of the resonance is reflected in the $K^0\Sigma^+$ Jackson frame angular distribution 
and was found to be $l\le 1$. Parity and angular momentum conservation then restrict the possible 
states to be $S_{11}$, $P_{11}$, $P_{13}$ ($N^*$) or $P_{33}$ ($\Delta^*$).  
The shape of proton angular distributions in the CM system for the $pp\to pB^*$ reaction were 
calculated under the assumption that one of these resonances is
involved. It turned out that all these calculated distributions have a relative angular momentum
in the $p-B^*$ system of $l\le 1$  in agreement with that 
deduced from the experimental proton CMS distribution. 
Considering that the angular distribution in the $K^0\Sigma^+$ Jackson frame is clearly anisotropic and
in view of the conclusions drawn from inspecting the distributions of invariant masses and helicity angles, 
$N(1710)P_{11}$ and/or $N(1720)P_{13}$ are the remaining 
candidate(s) for the dominating process of the $pp\to pK^0\Sigma^+$ reaction. Contributions from 
$N(1650)S_{11}$ and/or $\Delta(1600)P_{33}$ could also be present.

A simultaneous description of all 12 differential distributions measured was possible by assuming  
a single resonance  ($N(1720)$, $\Gamma = 150\,\mathrm{MeV/c^2}$) to dominate the reaction and 
applying weight functions (filters) on 
both the proton CMS angular distribution and the distribution in the $K^0\Sigma^+$ Jackson frame. 
These kinematic constraints are linearly independent in the four-dimensional reaction space. 
Hence, this simple approach to describe the reaction mechanism is self-consistent.

An advanced analysis of the data could be based on a partial wave analysis as 
performed by the Bonn-Gatchina group for various other reactions \cite{BonnGatchina}. 
This might stimulate the further development
of theoretical models which describe the associated strangeness production 
in proton-proton collisions from first principles.

\section*{Acknowledgment}
The authors would like to express their gratitude to the COSY staff for the operation 
of the accelerator during the experiments. 
This work was supported in part by grants from BMBF and Forschungszentrum J\"ulich (COSY-FFE).

% ----------------------------------------------------------------
% data tables start
% 

\begin{table*}
\caption{Differential cross sections  (in units of $n$b/$(9.0MeV/c^2)$) of the invariant mass distributions shown in fig.~\ref{fig:3059sigmaplusB} }
%Invariant mass data (Okt04, 3059, 801, 9, 0, KpS)
\label{tab:diffobsA}
     	\begin{tabular}{@{}rrrrrr}
  		\hline\noalign{\smallskip}
 $m_{K\Sigma^+}$ & $d\sigma/dm$ & $m_{Kp}$ & $d\sigma/dm$  & $m_{p\Sigma^+}$ &  $d\sigma/dm$ \\ 
  		\noalign{\smallskip}\hline\noalign{\smallskip}
  1674.50  &        --- \makebox[0.38cm]{}      &   1434.50  &        --- \makebox[0.38cm]{}      &   2129.50  &  $  31 \pm \makebox[0.56cm][r]{15.1} $ \\ 
  1683.50  &  $   6 \pm \makebox[0.56cm][r]{3.6} $ &  1443.50  &  $ 131 \pm \makebox[0.56cm][r]{36.1} $ &  2138.50  &  $  97 \pm \makebox[0.56cm][r]{28.0} $ \\ 
  1692.50  &  $ 164 \pm \makebox[0.56cm][r]{88.7} $ &  1452.50  &  $ 141 \pm \makebox[0.56cm][r]{24.6} $ &  2147.50  &  $ 130 \pm \makebox[0.56cm][r]{24.9} $ \\ 
  1701.50  &  $ 276 \pm \makebox[0.56cm][r]{29.2} $ &  1461.50  &  $ 248 \pm \makebox[0.56cm][r]{29.8} $ &  2156.50  &  $ 146 \pm \makebox[0.56cm][r]{24.8} $ \\ 
  1710.50  &  $ 361 \pm \makebox[0.56cm][r]{28.1} $ &  1470.50  &  $ 227 \pm \makebox[0.56cm][r]{26.5} $ &  2165.50  &  $ 175 \pm \makebox[0.56cm][r]{21.5} $ \\ 
  1719.50  &  $ 358 \pm \makebox[0.56cm][r]{28.3} $ &  1479.50  &  $ 310 \pm \makebox[0.56cm][r]{31.3} $ &  2174.50  &  $ 210 \pm \makebox[0.56cm][r]{25.1} $ \\ 
  1728.50  &  $ 389 \pm \makebox[0.56cm][r]{30.0} $ &  1488.50  &  $ 321 \pm \makebox[0.56cm][r]{29.7} $ &  2183.50  &  $ 253 \pm \makebox[0.56cm][r]{29.6} $ \\ 
  1737.50  &  $ 420 \pm \makebox[0.56cm][r]{32.8} $ &  1497.50  &  $ 305 \pm \makebox[0.56cm][r]{27.2} $ &  2192.50  &  $ 297 \pm \makebox[0.56cm][r]{29.1} $ \\ 
  1746.50  &  $ 394 \pm \makebox[0.56cm][r]{31.0} $ &  1506.50  &  $ 334 \pm \makebox[0.56cm][r]{28.2} $ &  2201.50  &  $ 301 \pm \makebox[0.56cm][r]{29.1} $ \\ 
  1755.50  &  $ 391 \pm \makebox[0.56cm][r]{34.2} $ &  1515.50  &  $ 380 \pm \makebox[0.56cm][r]{31.3} $ &  2210.50  &  $ 373 \pm \makebox[0.56cm][r]{29.6} $ \\ 
  1764.50  &  $ 391 \pm \makebox[0.56cm][r]{35.5} $ &  1524.50  &  $ 362 \pm \makebox[0.56cm][r]{28.7} $ &  2219.50  &  $ 315 \pm \makebox[0.56cm][r]{27.9} $ \\ 
  1773.50  &  $ 391 \pm \makebox[0.56cm][r]{36.9} $ &  1533.50  &  $ 345 \pm \makebox[0.56cm][r]{29.8} $ &  2228.50  &  $ 360 \pm \makebox[0.56cm][r]{30.9} $ \\ 
  1782.50  &  $ 326 \pm \makebox[0.56cm][r]{35.7} $ &  1542.50  &  $ 303 \pm \makebox[0.56cm][r]{26.2} $ &  2237.50  &  $ 392 \pm \makebox[0.56cm][r]{32.2} $ \\ 
  1791.50  &  $ 328 \pm \makebox[0.56cm][r]{36.9} $ &  1551.50  &  $ 301 \pm \makebox[0.56cm][r]{26.5} $ &  2246.50  &  $ 398 \pm \makebox[0.56cm][r]{31.0} $ \\ 
  1800.50  &  $ 279 \pm \makebox[0.56cm][r]{38.9} $ &  1560.50  &  $ 281 \pm \makebox[0.56cm][r]{27.6} $ &  2255.50  &  $ 353 \pm \makebox[0.56cm][r]{31.4} $ \\ 
  1809.50  &  $ 270 \pm \makebox[0.56cm][r]{45.2} $ &  1569.50  &  $ 244 \pm \makebox[0.56cm][r]{26.7} $ &  2264.50  &  $ 268 \pm \makebox[0.56cm][r]{24.1} $ \\ 
  1818.50  &  $ 163 \pm \makebox[0.56cm][r]{32.5} $ &  1578.50  &  $ 210 \pm \makebox[0.56cm][r]{29.6} $ &  2273.50  &  $ 185 \pm \makebox[0.56cm][r]{19.6} $ \\ 
  1827.50  &  $ 103 \pm \makebox[0.56cm][r]{28.3} $ &  1587.50  &  $ 163 \pm \makebox[0.56cm][r]{71.3} $ &  2282.50  &  $ 147 \pm \makebox[0.56cm][r]{17.6} $ \\ 
  1836.50  &  $  73 \pm \makebox[0.56cm][r]{40.9} $ &  1596.50  &  $  18 \pm \makebox[0.56cm][r]{9.5} $ &  2291.50  &     --- \makebox[0.38cm]{}    \\ 
  1845.50  &        --- \makebox[0.38cm]{}      &   1605.50  &        --- \makebox[0.38cm]{}      &   2300.50  &     --- \makebox[0.38cm]{}    \\ 
  	\noalign{\smallskip}\hline
	\end{tabular}
\end{table*}

\begin{table*}
\caption{Differential cross sections  (in units of $n$b/sr) of the angular distributions shown in fig.~\ref{fig:3059sigmaplusB} and \ref{fig:3059sigmaplusA}}
%Differential distributions. (Okt04, 3059, 801, 9, 0, KpS)
\label{tab:diffobsB}
     	\begin{tabular}{@{}rrrrrrrrrr}
  		\hline\noalign{\smallskip}
 cos & $\;\;\;\;\;\;\;\;\theta^*_p$ & $\;\;\;\;\;\;\;\;\theta^*_K$ & $\;\;\;\;\;\;\theta^*_{\Sigma^+}$  &  $\;\;\;\;\;\;\;\;\theta^{Rp{\Sigma^+}}_{bp}$ 	&   $\;\;\;\;\theta^{RKp}_{bK}$  &   $\;\;\;\;\theta^{R{\Sigma^+}K}_{b{\Sigma^+}}$&   $\;\;\;\;\theta^{Rp{\Sigma^+}}_{Kp}$ &   $\;\;\;\;\theta^{RKp}_{{\Sigma^+}K}$&   $\;\;\;\;\theta^{R{\Sigma^+}K}_{p{\Sigma^+}}$ \\ 
  		\noalign{\smallskip}\hline\noalign{\smallskip}
 \makebox[5mm][r]{ -0.95 }  &  $ 377 \pm \makebox[0.44cm][r]{68} $ 	&       ---    \makebox[0.38cm]{}   	&  $ 416 \pm \makebox[0.44cm][r]{122} $  & $ 354 \pm \makebox[0.44cm][r]{46} $ & $ 365 \pm \makebox[0.44cm][r]{94} $  & $ 430 \pm \makebox[0.44cm][r]{38} $    & $ 163 \pm \makebox[0.44cm][r]{65} $& $ 525 \pm \makebox[0.44cm][r]{57} $  & $ 238 \pm \makebox[0.44cm][r]{60} $    \\ 
 \makebox[5mm][r]{ -0.85 }  &  $ 368 \pm \makebox[0.44cm][r]{41} $ 	&       ---    \makebox[0.38cm]{}   	&  $ 351 \pm \makebox[0.44cm][r]{64} $   & $ 359 \pm \makebox[0.44cm][r]{36} $ & $ 344 \pm \makebox[0.44cm][r]{48} $  & $ 388 \pm \makebox[0.44cm][r]{36} $    & $ 107 \pm \makebox[0.44cm][r]{27} $& $ 470 \pm \makebox[0.44cm][r]{42} $  & $ 325 \pm \makebox[0.44cm][r]{38} $    \\ 
 \makebox[5mm][r]{ -0.75 }  &  $ 338 \pm \makebox[0.44cm][r]{28} $ 	& $ 319 \pm \makebox[0.44cm][r]{218} $ 	& $ 319 \pm \makebox[0.44cm][r]{53} $    & $ 288 \pm \makebox[0.44cm][r]{26} $ & $ 375 \pm \makebox[0.44cm][r]{31} $  & $ 351 \pm \makebox[0.44cm][r]{32} $    & $ 231 \pm \makebox[0.44cm][r]{44} $& $ 438 \pm \makebox[0.44cm][r]{39} $  & $ 368 \pm \makebox[0.44cm][r]{44} $    \\ 
 \makebox[5mm][r]{ -0.65 }  &  $ 359 \pm \makebox[0.44cm][r]{28} $ 	& $ 388 \pm \makebox[0.44cm][r]{161} $ 	& $ 345 \pm \makebox[0.44cm][r]{59} $    & $ 262 \pm \makebox[0.44cm][r]{24} $ & $ 331 \pm \makebox[0.44cm][r]{26} $  & $ 353 \pm \makebox[0.44cm][r]{31} $    & $ 298 \pm \makebox[0.44cm][r]{60} $& $ 430 \pm \makebox[0.44cm][r]{37} $  & $ 304 \pm \makebox[0.44cm][r]{34} $    \\ 
 \makebox[5mm][r]{ -0.55 }  &  $ 325 \pm \makebox[0.44cm][r]{25} $ 	& $ 290 \pm \makebox[0.44cm][r]{101} $ 	& $ 276 \pm \makebox[0.44cm][r]{46} $    & $ 291 \pm \makebox[0.44cm][r]{25} $ & $ 327 \pm \makebox[0.44cm][r]{25} $  & $ 278 \pm \makebox[0.44cm][r]{24} $    & $ 215 \pm \makebox[0.44cm][r]{39} $& $ 397 \pm \makebox[0.44cm][r]{36} $  & $ 346 \pm \makebox[0.44cm][r]{43} $    \\ 
 \makebox[5mm][r]{ -0.45 }  &  $ 316 \pm \makebox[0.44cm][r]{25} $ 	& $ 345 \pm \makebox[0.44cm][r]{96} $ 	& $ 281 \pm \makebox[0.44cm][r]{41} $    & $ 330 \pm \makebox[0.44cm][r]{26} $ & $ 329 \pm \makebox[0.44cm][r]{25} $  & $ 287 \pm \makebox[0.44cm][r]{25} $    & $ 273 \pm \makebox[0.44cm][r]{42} $& $ 323 \pm \makebox[0.44cm][r]{32} $  & $ 301 \pm \makebox[0.44cm][r]{40} $    \\ 
 \makebox[5mm][r]{ -0.35 }  &  $ 350 \pm \makebox[0.44cm][r]{29} $ 	& $ 281 \pm \makebox[0.44cm][r]{74} $ 	& $ 339 \pm \makebox[0.44cm][r]{40} $    & $ 278 \pm \makebox[0.44cm][r]{24} $ & $ 354 \pm \makebox[0.44cm][r]{26} $  & $ 314 \pm \makebox[0.44cm][r]{26} $    & $ 274 \pm \makebox[0.44cm][r]{44} $& $ 343 \pm \makebox[0.44cm][r]{35} $  & $ 349 \pm \makebox[0.44cm][r]{37} $    \\ 
 \makebox[5mm][r]{ -0.25 }  &  $ 310 \pm \makebox[0.44cm][r]{29} $ 	& $ 248 \pm \makebox[0.44cm][r]{48} $ 	& $ 273 \pm \makebox[0.44cm][r]{35} $    & $ 280 \pm \makebox[0.44cm][r]{24} $ & $ 282 \pm \makebox[0.44cm][r]{22} $  & $ 313 \pm \makebox[0.44cm][r]{25} $    & $ 280 \pm \makebox[0.44cm][r]{38} $& $ 305 \pm \makebox[0.44cm][r]{37} $  & $ 324 \pm \makebox[0.44cm][r]{36} $    \\ 
 \makebox[5mm][r]{ -0.15 }  &  $ 270 \pm \makebox[0.44cm][r]{29} $ 	& $ 314 \pm \makebox[0.44cm][r]{52} $ 	& $ 281 \pm \makebox[0.44cm][r]{34} $    & $ 310 \pm \makebox[0.44cm][r]{25} $ & $ 351 \pm \makebox[0.44cm][r]{27} $  & $ 293 \pm \makebox[0.44cm][r]{23} $    & $ 334 \pm \makebox[0.44cm][r]{42} $& $ 327 \pm \makebox[0.44cm][r]{38} $  & $ 274 \pm \makebox[0.44cm][r]{33} $    \\ 
 \makebox[5mm][r]{ -0.05 }  &  $ 304 \pm \makebox[0.44cm][r]{35} $ 	& $ 336 \pm \makebox[0.44cm][r]{45} $ 	& $ 310 \pm \makebox[0.44cm][r]{35} $    & $ 325 \pm \makebox[0.44cm][r]{27} $ & $ 310 \pm \makebox[0.44cm][r]{24} $  & $ 292 \pm \makebox[0.44cm][r]{22} $    & $ 320 \pm \makebox[0.44cm][r]{39} $& $ 354 \pm \makebox[0.44cm][r]{38} $  & $ 341 \pm \makebox[0.44cm][r]{35} $    \\ 
 \makebox[5mm][r]{ 0.05 }  	&  $ 345 \pm \makebox[0.44cm][r]{46} $ 	& $ 255 \pm \makebox[0.44cm][r]{33} $ 	& $ 352 \pm \makebox[0.44cm][r]{34} $    & $ 278 \pm \makebox[0.44cm][r]{25} $ & $ 271 \pm \makebox[0.44cm][r]{23} $  & $ 258 \pm \makebox[0.44cm][r]{21} $    & $ 359 \pm \makebox[0.44cm][r]{42} $& $ 343 \pm \makebox[0.44cm][r]{40} $  & $ 340 \pm \makebox[0.44cm][r]{35} $    \\ 
 \makebox[5mm][r]{ 0.15 }  	&  $ 323 \pm \makebox[0.44cm][r]{54} $ 	& $ 296 \pm \makebox[0.44cm][r]{32} $ 	& $ 314 \pm \makebox[0.44cm][r]{30} $    & $ 299 \pm \makebox[0.44cm][r]{24} $ & $ 246 \pm \makebox[0.44cm][r]{20} $  & $ 286 \pm \makebox[0.44cm][r]{23} $    & $ 354 \pm \makebox[0.44cm][r]{43} $& $ 262 \pm \makebox[0.44cm][r]{34} $  & $ 391 \pm \makebox[0.44cm][r]{40} $    \\ 
 \makebox[5mm][r]{ 0.25 }  	&  $ 280 \pm \makebox[0.44cm][r]{46} $ 	& $ 255 \pm \makebox[0.44cm][r]{27} $ 	& $ 300 \pm \makebox[0.44cm][r]{29} $    & $ 345 \pm \makebox[0.44cm][r]{28} $ & $ 270 \pm \makebox[0.44cm][r]{21} $  & $ 237 \pm \makebox[0.44cm][r]{20} $    & $ 344 \pm \makebox[0.44cm][r]{38} $& $ 311 \pm \makebox[0.44cm][r]{42} $  & $ 331 \pm \makebox[0.44cm][r]{41} $    \\ 
 \makebox[5mm][r]{ 0.35 }  	&  $ 339 \pm \makebox[0.44cm][r]{60} $ 	& $ 373 \pm \makebox[0.44cm][r]{33} $ 	& $ 328 \pm \makebox[0.44cm][r]{30} $    & $ 312 \pm \makebox[0.44cm][r]{25} $ & $ 263 \pm \makebox[0.44cm][r]{22} $  & $ 311 \pm \makebox[0.44cm][r]{25} $    & $ 396 \pm \makebox[0.44cm][r]{40} $& $ 273 \pm \makebox[0.44cm][r]{35} $  & $ 361 \pm \makebox[0.44cm][r]{36} $    \\ 
 \makebox[5mm][r]{ 0.45 }  	&  $ 275 \pm \makebox[0.44cm][r]{114} $ & $ 288 \pm \makebox[0.44cm][r]{26} $ 	& $ 288 \pm \makebox[0.44cm][r]{29} $    & $ 310 \pm \makebox[0.44cm][r]{25} $ & $ 326 \pm \makebox[0.44cm][r]{24} $  & $ 300 \pm \makebox[0.44cm][r]{24} $    & $ 412 \pm \makebox[0.44cm][r]{39} $& $ 286 \pm \makebox[0.44cm][r]{46} $  & $ 347 \pm \makebox[0.44cm][r]{36} $    \\ 
 \makebox[5mm][r]{ 0.55 }  	&  $ 419 \pm \makebox[0.44cm][r]{136} $ & $ 298 \pm \makebox[0.44cm][r]{27} $ 	& $ 366 \pm \makebox[0.44cm][r]{32} $    & $ 313 \pm \makebox[0.44cm][r]{26} $ & $ 315 \pm \makebox[0.44cm][r]{25} $  & $ 318 \pm \makebox[0.44cm][r]{27} $    & $ 400 \pm \makebox[0.44cm][r]{37} $& $ 247 \pm \makebox[0.44cm][r]{37} $  & $ 354 \pm \makebox[0.44cm][r]{37} $    \\ 
 \makebox[5mm][r]{ 0.65 }  	&  $ 351 \pm \makebox[0.44cm][r]{127} 	$ & $ 356 \pm \makebox[0.44cm][r]{29} $ & $ 350 \pm \makebox[0.44cm][r]{32} $    & $ 346 \pm \makebox[0.44cm][r]{29} $ & $ 302 \pm \makebox[0.44cm][r]{24} $  & $ 329 \pm \makebox[0.44cm][r]{29} $    & $ 394 \pm \makebox[0.44cm][r]{38} $& $ 221 \pm \makebox[0.44cm][r]{40} $  & $ 363 \pm \makebox[0.44cm][r]{37} $    \\ 
 \makebox[5mm][r]{ 0.75 }  	&  $ 350 \pm \makebox[0.44cm][r]{97} $ 	& $ 331 \pm \makebox[0.44cm][r]{28} $ 	& $ 340 \pm \makebox[0.44cm][r]{36} $    & $ 348 \pm \makebox[0.44cm][r]{28} $ & $ 354 \pm \makebox[0.44cm][r]{30} $  & $ 353 \pm \makebox[0.44cm][r]{31} $    & $ 464 \pm \makebox[0.44cm][r]{38} $& $ 297 \pm \makebox[0.44cm][r]{53} $  & $ 361 \pm \makebox[0.44cm][r]{42} $    \\ 
 \makebox[5mm][r]{ 0.85 }  	&  $ 240 \pm \makebox[0.44cm][r]{139} $ & $ 373 \pm \makebox[0.44cm][r]{31} $ 	& $ 405 \pm \makebox[0.44cm][r]{78} $    & $ 404 \pm \makebox[0.44cm][r]{33} $ & $ 391 \pm \makebox[0.44cm][r]{56} $  & $ 409 \pm \makebox[0.44cm][r]{42} $    & $ 453 \pm \makebox[0.44cm][r]{39} $& $ 245 \pm \makebox[0.44cm][r]{51} $  & $ 296 \pm \makebox[0.44cm][r]{40} $    \\ 
 \makebox[5mm][r]{ 0.95 }  	&        ---   \makebox[0.38cm]{}    	&  $ 451 \pm \makebox[0.44cm][r]{36} $ 	& $ 328 \pm \makebox[0.44cm][r]{148} $   & $ 421 \pm \makebox[0.44cm][r]{34} $ & $ 285 \pm \makebox[0.44cm][r]{89} $  & $ 378 \pm \makebox[0.44cm][r]{37} $    & $ 537 \pm \makebox[0.44cm][r]{40} $& $ 194 \pm \makebox[0.44cm][r]{48} $  & $ 298 \pm \makebox[0.44cm][r]{66} $    \\ 
  	\noalign{\smallskip}\hline
	\end{tabular}
\end{table*}

%
% data tables end
%-----------------------------------------------------------------

%
% BibTeX users please use
% \bibliographystyle{}
% \bibliography{}

\appendix
\section{Datatables}

\end{document}